\documentclass[aps,pra,twocolumn,groupedaddress,showpacs,superscriptaddress]{revtex4-1}
\usepackage{amssymb,amsmath,graphicx,epstopdf,hyperref,amsthm}
\begin{document}
\interfootnotelinepenalty=10000

\title{Bloch sphere colourings and Bell inequalities}

\author{Adrian Kent}
\affiliation{Centre for Quantum Information and Foundations, DAMTP, Centre for
Mathematical Sciences,
University of Cambridge, Wilberforce Road, Cambridge, CB3 0WA, United Kingdom}
\affiliation{Perimeter Institute for Theoretical Physics, 31 Caroline Street
North, Waterloo, Ontario, Canada N2L 2Y5}
\author{Dami\'an Pital\'ua-Garc\'ia}

\affiliation{Centre for Quantum Information and Foundations, DAMTP, Centre for
Mathematical Sciences,
University of Cambridge, Wilberforce Road, Cambridge, CB3 0WA, United Kingdom}


\begin{abstract}
We consider the quantum and local hidden variable (LHV) correlations obtained by measuring a pair of qubits
by projections defined by randomly chosen axes separated by an
angle $\theta$. LHVs predict binary colourings of the Bloch
sphere with antipodal points oppositely coloured.
We prove Bell inequalities separating the LHV predictions from the singlet quantum correlations
for $\theta\in\bigl(0,\frac{\pi}{3}\bigr)$.
We raise and explore the hypothesis that, for a continuous
range of $\theta>0$, the maximum LHV anticorrelation is obtained by
assigning to each qubit a colouring with one
hemisphere black and the other white.
\end{abstract}


\maketitle

\section{Introduction}
According to quantum theory, space-like separated
experiments performed on entangled particles can produce outcomes
whose correlations violate Bell inequalities \cite{Bell} that
would be satisfied if the experiments could be described by local
hidden variable theories (LHVTs). Many experiments have tested the quantum
prediction of
nonlocal causality
(e.g.
\cite{ADR82,WJSWZ81,TBZG98,*GZ99,RKMSIMW01,MMMOM08,SBHGZ08,GMRWKBLCGNUZ13}).
The observed violations of Bell inequalities are consistent with
quantum theory.   They refute LHVTs with
overwhelmingly
high degrees of confidence, modulo some known loopholes
that arise from the difficulty in carrying out theoretically ideal
experiments -- most notably the locality loophole (closed in \cite{ADR82,WJSWZ81,TBZG98,GZ99}),
the detection efficiency loophole (\cite{Pearle70}, closed in
\cite{RKMSIMW01,MMMOM08,GMRWKBLCGNUZ13})
and the collapse locality loophole (\cite{K05}, addressed in
\cite{SBHGZ08},
though not fully closed). 

Typically, Bell experiments test the CHSH inequality
\cite{CHSH69} in an EPR-Bohm experiment
\cite{EPR35,Bohmbook} in which two
entangled particles are sent to different experimental setups at different
locations. One setup is controlled by Alice, who performs one
of two possible measurements $A\in\lbrace0,1\rbrace$;
the other by Bob, who similarly performs $B\in\lbrace0,1\rbrace$.
Alice's and Bob's outcomes $a$ and $b$
are assigned numerical values $a,b\in\lbrace1,-1\rbrace$,
corresponding to `spin up' or `spin down' for spin measurements about
given axes on spin-$\frac{1}{2}$ particles.  The experiments must be completed
at space-like separated regions. The experiment is repeated many
times, ideally under identical experimental conditions.
We define the correlation $C(A,B)$ as the average value of the product
of Alice's and Bob's outcomes in experiments where measurements $A$
and $B$ are chosen.

According to deterministic LHVTs, the outcomes $a$, $b$ are
determined respectively by the measurement choices $A$, $B$ and by
hidden variables $\lambda$ shared by both particles.
Thus, $a=a(A,\lambda), b=b(B,\lambda)$.
An LHVT also assigns a probability distribution $\rho(\lambda)$,
independent of $A$ and $B$, to the hidden variables, satisfying
$\rho(\lambda)\geq 0$ and $\int_{\Lambda} d\lambda \rho(\lambda)=1$,
where $\Lambda$ is the sample space of hidden variables.
Probabilistic LHVTs can be described by the same equations,
extending the definitions of $\lambda$ and $\rho$ to allow for probabilistic
measurement outcomes; we can thus focus on deterministic LHVTs without
loss of generality.
An LHVT predicts $C(A,B)=\int_{\Lambda} d\lambda
\rho(\lambda)a(A,\lambda)b(B,\lambda)$.
Such correlations satisfy the CHSH inequality \cite{CHSH69}: $ I_2 = \bigl\lvert C(0,0)+C(1,1)+C(1,0)-C(0,1)\bigr\rvert \leq 2 \, .$

Consider for definiteness the EPR-Bohm experiment performed on
spin-$\frac{1}{2}$ particles in the singlet state
$\lvert\Psi^-\rangle=\frac{1}{\sqrt{2}}\bigl(\lvert \uparrow\rangle\lvert
\downarrow\rangle-\lvert
\downarrow\rangle\lvert \uparrow\rangle\bigr)$. Alice and Bob measure their
particle spin projection along the directions $\vec{a}_A$ and $\vec{b}_B$, respectively.
As before, Alice and Bob choose a measurement
from a set of two elements, that is, $A, B\in\lbrace0,1\rbrace$.
In general, the vectors $\vec{a}_A$ and $\vec{b}_B$ can point along
any direction in three-dimensional Euclidean space, and the sets of
their possible values define Bloch spheres $\mathbb{S}^2$.
The correlation predicted by quantum theory is $Q(\theta) = -\cos \theta$,
where $\cos \theta = \vec{a}_A \cdot \vec{b}_B$.
Sets of measurement axes can be found for which the quantum
correlations violate the CHSH inequality, $I_2^{\text{QM}} > 2$, up to the Cirel'son
\cite{C80} bound $I_2^{\text{QM}} \leq 2\sqrt{2}$.

When Alice's and Bob's measurement choices belong to a set of $N$ possible
elements, the correlations predicted by LHVTs satisfy the Braunstein-Caves
inequality \cite{BC90}:
\begin{equation}
\label{eq:7}
I_N=\Biggl\lvert
\sum_{k=0}^{N-1}\!\!C(k,k)+\sum_{k=0}^{N-2}\!\!C(k+1,k)-C(0,N-1)
\Biggr\rvert\leq 2N-2.
\end{equation}

The CHSH inequality is a special case of the Braunstein-Caves
inequality with $N=2$.   We are interested here in exploring Bell
inequalities that generalize the CHSH and Braunstein-Caves
inequalities, in the following sense.
Instead of restricting Alice's and Bob's measurement choices
to a finite set, we allow them to choose any spin measurement
axes,  $\vec{a}$
and $\vec{b}$.   However, we constrain these axes
to be separated by a fixed angle $\theta$,
so $\cos \theta = \vec{a} \cdot \vec{b}$.
The maximal violation of the Braunstein-Caves inequality by quantum correlations, given by $I_N^{\text{QM}} = 2N\cos\bigl(\frac{\pi}{2N}\bigr)$ \cite{W06}, arises
for fixed sets of pairs of axis choices that satisfy this
constraint with $ \theta = \frac{\pi}{2N}$.  We consider
experiments where pairs of axes separated by $\theta$ are chosen
randomly and where $\theta$ is unrestricted.     Our work contributes
to understanding how to quantify quantum nonlocality, by studying
a natural class of Bell inequalities that has not previously been considered.
As well as proving new inequalities, our work raises new questions and 
suggests new techniques that we hope will be developed further.   

Another more practical motivation is to
explore simple Bell tests that might allow quantum theory and LHVT
to be distinguished somewhat more efficiently, particularly in the
adversarial context of quantum cryptography.
Here an eavesdropper and/or malicious device manufacturer
may be trying to spoof the correlations of 
a singlet using locally held or
generated information.   
Of course, given sufficient guarantees about the devices
involved, modulo the loopholes mentioned above, and with
sufficiently many runs, any Bell test can expose such spoofing.
However, in practical situations in which the number of possible tests
is limited, users would like to ensure that such eavesdropping
attacks can be detected as efficiently as possible.   
Standard CHSH tests simplify the eavesdropper's problem, 
by informing her in advance that she need only generate
outcomes for a small set of possible measurements.     
By comparison, tests involving randomly chosen axes
give the eavesdropper no such information \footnote{
One possibility here is for Alice and Bob to fix in advance the
value of $\theta$ and a list of random pairs of axes separated
by $\theta$.      Another would be to make random independent 
choices and then generate plots of the
correlations as a function of $\theta$.   
This second type of test would be generated automatically by  
quantum key distribution schemes that require 
Alice and Bob to make completely random measurements
on each qubit (e.g. \cite{KBMS11}).}.  
A first step towards understanding Alice's and Bob's optimal 
test strategy in such contexts is to identify the
full range of Bell inequalities available.  
 
\section{Bloch sphere colourings and correlation functions}
We explore LHVTs in which Alice's and Bob's spin measurement results are
given by $a(\vec{a},\lambda)$ and $b\bigl(\vec{b},\lambda\bigr)$,
respectively; where $\lambda$ is a local hidden variable common to
both particles. For fixed $\lambda$, we can describe the
functions $a$ and $b$ by two binary (black and white) colourings of spheres,
associated with $a$ and $b$, respectively, where black (white)
represents the outcome `1'~(`-1'). Different sphere colourings
are associated with different values of $\lambda$. 
To look at specific cases, we drop the
$\lambda$-dependence and include a label $x$ that indicates a
particular pair of colouring functions $a_x(\vec{a})$ and
$b_x(\vec{b}\!~)$.

Measuring spin along $\vec{a}$ with outcome $1~(-1)$ is equivalent to
measuring spin along $-\vec{a}$
with outcome $-1~(1)$.   The colouring functions $a$ and $b$ 
defining any LHVT are thus necessarily \emph{antipodal}
functions:
\begin{equation}
\label{eq:antipodalx}
a_x(\vec{a})=-a_x(-\vec{a}),\qquad
b_x\bigl(\vec{b}~\!\bigr)=-b_x\bigl(-\vec{b}~\!\bigr),
\end{equation}
for all $\vec{a},\vec{b}\in\mathbb{S}^2$.

We notice that the antipodal property
arises due to the definition of a dichotomic measurement on the sphere for arbitrary deterministic LHVTs.
For an arbitrary probabilistic theory, this property would read
\begin{equation}
\label{eq:antipodalprob}
P_x\bigl(\mu a,\nu b\vert \mu\vec{a},\nu\vec{b}~\!\bigr)=P_x\bigl(a,b\vert \vec{a},\vec{b}~\!\bigr),
\end{equation}
where $\mu,\nu,a,b\in\{\pm1\}$, $\vec{a},\vec{b}\in\mathbb{S}^2$ and the label $x$ indicates a particular probabilistic theory being considered. Equation (\ref{eq:antipodalprob}) holds because a measurement is defined by a pair of opposite axes, $\vec{a}$ and $-\vec{a}$, and inverting their sense corresponds only to relabelling the measurement outcomes.

We define $\mathcal{X}$ as
the set of all colourings $x$ satisfying the antipodal property,
Eq.~(\ref{eq:antipodalx}). For example, a simple colouring of the
spheres satisfying the antipodal property is \emph{colouring 1}, in
which, for one sphere, one hemisphere is completely black and the
other one is completely white, and the colouring is reversed for the
other sphere (see Fig.~\ref{fig2}).  

The correlation for outcomes of measurements about randomly chosen axes
separated by $\theta$ for the pair of colouring
functions labelled by $x$ is
\begin{equation}
\label{eq:14}
C_x(\theta)=\frac{1}{8\pi^2}\int_{\mathbb{S}^2}dA
a_x(\vec{a})\int_{0}^{2\pi}d\omega b_x\bigl(\vec{b}\!~\bigr),
\end{equation}
where $dA$ is the area element of the sphere corresponding to Alice's
axis $\vec{a}$ and $\omega$ is an angle in the range $[0,2\pi]$ along
the circle described by Bob's axis $\vec{b}$ with an angle $\theta$
with respect to $\vec{a}$ . A general correlation is of the
form $C(\theta)=\int_{\mathcal{X}} dx\mu (x) C_x(\theta)$, where
$\mu(x)$ is 
a probability
distribution over
$\mathcal{X}$.

If all colourings $x\in\mathcal{X}$ satisfy
$Q_{\rho_{\text{L}}}(\theta)< C^{\text{L}} ( \theta )\leq C_x(\theta)$
or $C_x(\theta)\leq C^{\text{U}} ( \theta ) <
Q_{{\rho}_{\text{U}}}(\theta)$ for quantum correlations
$Q_{\rho_{\text{L}}}(\theta)$ and $Q_{{\rho}_{\text{U}}}(\theta)$
obtained with particular two-qubit states ${\rho_{\text{U}}}$ and
${\rho_{\text{L}}}$, and some identifiable lower and upper bounds,
$C^{\text{L}} ( \theta )$ and $C^{\text{U}} ( \theta )$, respectively,
then a general correlation $C(\theta)$ must satisfy the same
inequalities. Our aim here is to explore this possibility via
intuitive arguments and numerical and analytic results. We focus on
the case ${\rho_{\text{L}}}=\lvert\Psi^-\rangle\langle\Psi^-\rvert$,
for which $Q_{\rho_{\text{L}}}(\theta)\equiv Q(\theta)=-\cos\theta$,
which is the maximum quantum anticorrelation for a given angle
$\theta$ (see Sec. \ref{sec:RQE} for details and
related questions).  We begin with some suggestive observations.

First, we consider colouring functions $x\in\mathcal{X}$ for which the
probability that Alice and Bob obtain opposite outcomes when they
choose the same measurement, averaged uniformly over all measurement
choices, is
\begin{equation}
\label{eq:17.1}
P(a_x=-b_x\vert\theta=0)=1-\gamma.
\end{equation}
In general, $0\leq\gamma\leq 1$. We first consider small values of $\gamma$
and seek Bell inequalities distinguishing
quantum correlations for the singlet from classical correlations
for which an anticorrelation is observed with probability $1-\gamma$ when the
same measurement axis is chosen on both sides.
Experimentally, we can verify quantum nonlocality using these results
if we carry out nonlocality tests that include some
frequency of anticorrelation tests about a
randomly chosen axis (chosen independently for each test).
The anticorrelation tests allow statistical bounds on $\gamma$,
which imply statistical tests of nonlocality via
the $\gamma$-dependent Bell inequalities.

In the limiting case $\gamma=0$, we have
\begin{equation}
\label{eq:17}
a_x\left(\vec{a}\right)=-b_x\left(\vec{a}\right),
\end{equation}
for all $\vec{a}\in\mathbb{S}^2$. 
This case is quite interesting theoretically, in that
one might hope to prove stronger results assuming perfect
anticorrelation.  We describe some numerical explorations of
this case below.

Second, for any pair of colourings $x\in\mathcal{X}$ and
$\theta\in[0,\pi]$, we have $C_x(\pi-\theta)=-C_x(\theta)$. This can
be seen as follows. For a fixed $\vec{a}$, the circle with angle
$\theta=\theta'$ around the axis $\vec{a}$, defined by the angle
$\omega$ in Eq.~(\ref{eq:14}) contains a point $\vec{b}$ that is
antipodal to a point on the circle with angle $\theta=\pi-\theta'$
around $\vec{a}$. Since the colouring is antipodal, we have that the
value of the integral $\int_0^{2\pi}d\omega b_x\bigl(\vec{b}\!~\bigr)$
in Eq.~(\ref{eq:14}) for $\theta=\theta'$ is the negative of the
corresponding integral for $\theta=\pi-\theta'$. It follows that
$C_x(\pi-\theta')=-C_x(\theta')$. Therefore, in the rest of this
paper, we restrict to consider correlations for the range
$\theta\in\bigl[0,\frac{\pi}{2}\bigr]$, unless otherwise stated. From
the previous argument, we have
$C_x\bigl(\frac{\pi}{2}\bigr)=-C_x\bigl(\frac{\pi}{2}\bigr)$, which
implies that $C_x\bigl(\frac{\pi}{2}\bigr)=0$. We also have that
$C_x(0)=1-2P(a_x=-b_x\vert\theta=0)$, so the LHVTs we consider
give $C_x(0)=-1+2\gamma$. The LHV correlations given by
Eqs.~(\ref{eq:14}) and (\ref{eq:17.1}) in the case $\gamma=0$ thus
coincide with the singlet-state quantum correlations for $\theta=0$
and $\theta=\frac{\pi}{2}$, where $Q(0) = C_x (0) =-1$ and
$Q\bigl(\frac{\pi}{2}\bigr)= C_x \bigl( \frac{\pi}{2} \bigr) = 0$.

Third, consider colouring 1, defined above.  We have
$C_1(\theta)=-\bigl(1-\frac{2\theta}{\pi}\bigr)$, for
$\theta\in\bigl[0,\frac{\pi}{2}\bigr]$. This is easily seen as
follows. For any two different points on the spheres defining
colouring 1, $\vec{a}$ in one sphere and $\vec{b}$ in the oppositely
coloured one, an arc of angle $\pi$ of the great circle passing
through $\vec{a}$ and $\vec{b}$ is completely black and the other arc
of angle $\pi$ is completely white. Thus, given that the pair of
vectors $\vec{a}$ and $\vec{b}$ are chosen randomly, subject to the
constraint of angle separation $\theta$, the probability that both
$\vec{a}$ and $\vec{b}$ are in oppositely coloured regions is
$P(a_1=-b_1\vert\theta)=\frac{\pi-\theta}{\pi}=1-\frac{\theta}{\pi}$. Thus,
the correlation for colouring 1 is
$C_1(\theta)=1-2P(a_1=-b_1\vert\theta)=-1+\frac{2\theta}{\pi}$. That
is, $C_1 ( \theta )$ linearly interpolates between the values at $C_1
( 0 ) = -1 $, which is common to all colourings with $\gamma=0$, and $
C_1\bigl( \frac{\pi}{2} \bigr) = 0$, which is common to all
colourings, and we have $0 > C_1(\theta) > Q(\theta)$ for
$\theta\in\bigl(0,\frac{\pi}{2}\bigr)$.

Then, in the following section we present some lemmas and a theorem.

\section{Hemispherical Colouring Maximality Hypotheses}

In this section, we motivate two \emph{hemispherical colouring
  maximality
hypotheses}.  These make precise the intuition that,
for a continuous range of $\theta >0$, the maximum LHV anticorrelation is obtained by colouring $1$.

We first consider the following lemmas, whose proofs are given in Appendix \ref{appendix proofs}.

\newtheorem{lemma1}{Lemma}
\begin{lemma1}
\label{lemma1}
For any colouring $x\in\mathcal{X}$ satisfying Eq.~(\ref{eq:17.1}) and any
$\theta\in\bigl(0,\frac{2\pi}{3}\bigr] $, we have $-1+\frac{2}{3}\gamma\leq
C_x(\theta)\leq\frac{1}{3}+\frac{2}{3}\gamma$.
\end{lemma1}

\theoremstyle{remark}
\newtheorem{remark1}{Remark}
\begin{remark1}
\label{remark1}
Unsurprisingly, since small $\gamma$ implies near-perfect
anticorrelation
at $\theta=0$, we see that for $\theta\in\bigl(0,\frac{2\pi}{3}\bigr] $ and
$\gamma$ small there are
no colourings with very strong correlations.  However,
strong anticorrelations are possible for small $\theta$.
We are interested in bounding these.
\end{remark1}

\theoremstyle{plain}
\newtheorem{lemma2}[lemma1]{Lemma}
\begin{lemma2}
\label{lemma2}
For any colouring $x\in\mathcal{X}$ satisfying Eq.~(\ref{eq:17.1}), any integer
$N>2$ and any
$\theta\in\bigl[\frac{\pi}{N},\frac{\pi}{N-1}\bigr)$,
we have $C_x ( \theta ) \geq C_1\bigl( \frac{\pi}{N} \bigr)-2\gamma$.
\end{lemma2}

\theoremstyle{remark}
\newtheorem{remark2}[remark1]{Remark}
\begin{remark2}
\label{remark2}
In other words, for small $\theta$,
$C_1 (\theta )$ is very close to the maximal possible anticorrelation
for LHVTs when $\gamma \ll \theta$.

\end{remark2}

Geometric intuitions also suggest bounds on $C_x (\theta )$ that
are maximised by colouring $1$ for small $\theta$. 
Consider {\it simple colourings}, in which
a set of (not necessarily connected) piecewise differentiable curves of finite
total length separate black and white regions.  (Points lying on these
curves may have either colour.)
Intuition suggests that, for small $\theta$ and
simple colourings with $\gamma=0$, the quantity $1 +C_x(\theta)$, which
measures the deviation from pure anticorrelation,
should be bounded by a quantity roughly proportional to the length of the
boundary between the black
and white areas of the sphere colouring $x\in\mathcal{X}$.
Since colouring $1$ has the smallest such boundary (the equator),
this might suggest that
$ C_x (\theta ) \geq  C_1 (\theta
)\, , $ for small $\theta$ and for all simple colourings
$x\in\mathcal{X}$ with $\gamma=0$.   Intuition also suggests that any
non-simple colouring will
produce less anticorrelation than the optimal simple colouring, because
regions in which black and white colours alternate with arbitrarily
small separation tend to wash out anticorrelation.
These intuitive arguments are clearly not rigorous
as currently formulated.  For example, they ignore the
possibility of sequences of colourings $C_i (\theta )$
and angles $\theta_i \rightarrow 0$
such that $C_i (\theta_i ) < C_1 (\theta_i )$, while
$\lim_{\theta \rightarrow 0 } ( C_i (\theta ) - C_1 ( \theta ) )> 0$
for all $i$ (see \cite{DPGthesis} for an extended discussion).   Still, they are suggestive, at least in generating
hypotheses to be investigated.

These various observations motivate us to explore what we call
the weak hemispherical colouring maximality hypothesis (WHCMH).

\theoremstyle{plain}
\newtheorem*{WHCMH}{The WHCMH}
\begin{WHCMH}
There exists an angle
$\theta_{\text{max}}^{\text{w}}\in\bigl(0,\frac{\pi}{2}\bigr)$ such
that for
every colouring $x\in\mathcal{X}$ with $\gamma=0$ and every angle
$\theta\in[0,\theta_{\text{max}}^{\text{w}}]$, $C_x(\theta)\geq C_1(\theta)$.
\end{WHCMH}

The WHCMH considers models with perfect anticorrelation for $\theta=0$, because we are
interested in
distinguishing LHV models from the quantum singlet state, which
produces perfect anticorrelations for $\theta=0$.
Of course, there is a symmetry in the space of LHV models given by
exchanging the colours of one qubit's sphere, which maps
$\gamma \rightarrow 1 - \gamma$ and $C_x ( \theta ) \rightarrow - C_x (
\theta)$.  The WHCMH thus also implies that $ C_x ( \theta ) \leq - C_1 (\theta
)$
for all colourings $x\in\mathcal{X}$ with $\gamma=1$.

It is also interesting to investigate stronger versions of the WHCMH
and related questions.
For instance, is it the case that for every angle
$\theta\in\bigl(\theta_{\text{max}}^{\text{w}},\frac{\pi}{2}\bigr)$
there exists a colouring
$x'\in\mathcal{X}$ with $\gamma=0$ such that $C_{x'}(\theta)<C_1(\theta)$?
Further, does this hypothesis still hold true (not necessarily for the same
$\theta_{\text{max}}^{\text{w}}$) if we consider general local hidden variable
models corresponding to independently chosen colourings for the
two qubits, not constrained by any choice of the correlation parameter
$\gamma$? 

The following theorem and lemmas, whose proofs are presented in Appendix \ref{appendix proofs},
give some relevant bounds.

\newtheorem{theorem1}{Theorem}
\begin{theorem1}
\label{theorem1}
For any colouring $x\in\mathcal{X}$, any integer $N\geq 2$ and any
$\theta\in\bigl[\frac{\pi}{2N},\frac{\pi}{2(N-1)}\bigr)$,
we have $ C_1\bigl( \frac{\pi}{2N} \bigr)\leq C_x ( \theta ) \leq -C_1\bigl(
\frac{\pi}{2N} \bigr)$.
\end{theorem1}

\theoremstyle{remark}
\newtheorem{remark3}[remark1]{Remark}
\begin{remark3}
\label{remark3}
In particular, for small $\theta$,
$-C_1 (\theta )$ and $C_1 (\theta )$ are very close to the maximal possible
correlation and anticorrelation for any LHVT, respectively.
\end{remark3}

\theoremstyle{plain}
\newtheorem{lemma3}[lemma1]{Lemma}
\begin{lemma3}
\label{lemma3}
If any colouring $x\in\mathcal{X}$ obeys $C_x ( \theta) < C_1 (\theta
) ~\Bigl(C_x ( \theta) > -C_1 (\theta)\Bigr)$ for some
$\theta\in\bigl(\frac{\pi}{M+1},\frac{\pi}{M}\bigr] $ and an integer $M\geq 2$
then there are angles $\theta_j\equiv \frac{\pi}{M+1-j}-\theta$ with
$j=1,2,\ldots,M-1$, which satisfy $0\leq\theta_j<\theta$ if $j<\frac{M}{2}+1$,
and $\frac{\pi}{2}>\theta_j>\theta$ if $j\geq \frac{M}{2}+1$, such that $C_x (
\theta_j) > C_1 (\theta_j) ~\Bigl(C_x ( \theta_j) < -C_1 (\theta_j )\Bigr)$.
\end{lemma3}

\theoremstyle{remark}
\newtheorem{remark4}[remark1]{Remark}
\begin{remark4}
\label{remark4}
In this sense (at least), the
anticorrelations defined by $C_1$ and the correlations defined by $-C_1$
cannot be dominated by
any other colourings.
\end{remark4}

\theoremstyle{plain}
\newtheorem{lemma4}[lemma1]{Lemma}
\begin{lemma4}
\label{lemma4}
For any colouring $x\in\mathcal{X}$ and any
$\theta\in\bigl(0,\frac{\pi}{3}\bigr)$,
we have $ Q ( \theta )<C_x ( \theta )<-Q ( \theta )$.
\end{lemma4}

\theoremstyle{remark}
\newtheorem{remark5}[remark1]{Remark}
\begin{remark5}
\label{remark5}
This inequality separates all possible
LHV correlations $C_x(\theta)$ from the singlet-state quantum
correlations $Q(\theta)$ for all
$\theta\in\bigl(0,\frac{\pi}{3}\bigr)$.

\end{remark5}

The previous observations motivate the strong
hemispherical colouring maximality hypothesis (SHCMH).

\theoremstyle{plain}
\newtheorem*{SHCMH}{The SHCMH}
\begin{SHCMH}
There exists an angle
$\theta_{\text{max}}^{\text{s}}\in\bigl(0,\frac{\pi}{2}\bigr)$ such
that for
every colouring $x\in\mathcal{X}$ and every angle
$\theta\in[0,\theta_{\text{max}}^{\text{s}}]$, $C_1(\theta)\leq C_x(\theta)\leq
-C_1(\theta)$.
\end{SHCMH}

Note that the SHCMH applies
to all colourings, without any assumption of perfect anticorrelation for $\theta=0$.
If the SHCMH is true then so is the WHCMH.
In this case, we have that $\theta_{\text{max}}^{\text{s}}\leq
\theta_{\text{max}}^{\text{w}}$. Thus, an upper bound on
$\theta_{\text{max}}^{\text{w}}$ implies an upper bound on
$\theta_{\text{max}}^{\text{s}}$.

\section{Numerical results}

We investigated the WHCMH numerically by computing
the correlation $C_x(\theta)$ for various colouring
functions that satisfy the antipodal property, Eq.~(\ref{eq:antipodalx}), the
condition
(\ref{eq:17}), and
that have azimuthal symmetry (see Fig.~\ref{fig2}). Details of our numerical work are
given in Appendix \ref{appendix nr}. Our numerical results
are consistent with the WHCMH for $\theta_{\text{max}}^{\text{w}} \leq 0.386
\pi$, and with the SHCMH for $\theta_{\text{max}}^{\text{s}} \leq 0.375
\pi$, but do not give strong evidence for these values.   Nor do the numerical
results, {\it per se}, constitute compelling evidence for the
WHCMH and SHCMH, although they confirm that the underlying
intuitions hold for some simple colourings. 

We note that the slightly improved bound
$\theta_{\text{max}}^{\text{s}}\leq 0.345\pi$ was obtained in
\cite{DPGthesis}. Further details are given in Appendix \ref{appendix nr}.

\begin{figure}
\includegraphics{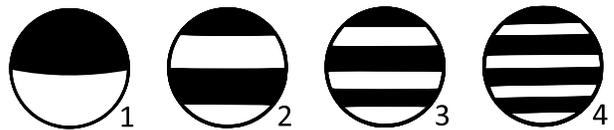}
 \caption{\label{fig2}Some antipodal colouring functions $a_x$ on the
   sphere, see  Appendix \ref{appendix nr} for definitions.
Their correlations $C_x(\theta)$, computed from Eq.~(\ref{eq:14}), subject to
the constraint (\ref{eq:17}), are plotted
in Appendix \ref{appendix nr}.}
\end{figure}

\section{Related questions for exploration}
\label{sec:RQE}
An interesting related question is, for an arbitrary two-qubit state $\rho$ and qubit projective
measurements performed by Alice and Bob corresponding to random Bloch vectors
separated by an angle $\theta$, what are the maximum values of the quantum correlations and anticorrelations $Q_\rho(\theta)$, and which states
achieve them? We show that the maximum quantum anticorrelations and correlations are $Q_\rho(\theta)=-\cos\theta$, achieved by the singlet state $\rho=\lvert\Psi^-\rangle\langle\Psi^-\rvert$, and $Q_\rho(\theta)=\frac{1}{3}\cos\theta$, achieved by the other Bell states, $\rho=\lvert\Phi^\pm\rangle\langle\Phi^\pm\rvert$ and $\rho=\lvert\Psi^+\rangle\langle\Psi^+\rvert$, respectively. This result follows because, as we show in Appendix \ref{appendix questions}, we have
\begin{equation}
\label{eq:new}
-\cos{\theta}\leq Q_\rho(\theta) \leq \frac{1}{3}\cos{\theta}.
\end{equation}

Another related question that we do not explore further here is, for a
fixed given angle $\theta$ separating Alice's and Bob's measurement
axes, what are the maximum correlations and anticorrelations, if in
addition to the two-qubit state $\rho$, Alice and Bob have other
resources? For example, Alice and Bob could have an arbitrary
entangled state on which they perform arbitrary local quantum
operations and measurements. In a different scenario, Alice and Bob
could have some amount of classical or quantum communication.
Another possibility is for Alice and Bob to share arbitrary
no-signalling resources, not necessarily quantum, with no
communication 
allowed. Different variations of the
task described above with continuous parameters can be
investigated.

One might ask what constraints the no-signalling
principle places on the correlations and
anticorrelations.   
A generalised 
PR-box \cite{PR94} gives the correlation $C ( \theta ) ={\rm sign}( \pi/2 -
\theta )$, which in one natural sense defines the strongest
correlations consistent with Eq.~(\ref{eq:antipodalprob}). 
Another relevant observation is that the antipodal property
(\ref{eq:antipodalprob}), expressed in the equivalent form
$C(\pi-\theta)=-C(\theta)$, together with a continuity assumption,
implies that quantum nonlocal correlations are not dominated
\cite{K13.2}: If a correlation $C(\theta)$ produces a violation of the
CHSH inequality stronger than the violation given by the singlet-state
quantum correlation $Q(\theta)$ for a given set of measurement axes
then there exists another set of measurement axes for which
$C(\theta)$ gives a violation (or none) that is weaker than the
violation given by $Q(\theta)$.
It would be interesting to clarify further the relationship between
measures of nonlocality, including those investigated here, and 
no-signalling.   

Other related questions are given in Appendix \ref{appendix questions}.

\section{Discussion}

We have explored here what can be learned by carrying out local
projective measurements about completely randomly chosen
axes, separated by an angle $\theta$, on a pair of qubits.
This is not currently a standard way of testing for
entanglement or nonlocality, but we have shown that
it distinguishes quantum correlations from those predicted
by local hidden variables for a wide range of $\theta$.
In particular, we find Bell inequalities for
$\theta\in\bigl(0,\frac{\pi}{2}\bigr)$, given by Theorem \ref{theorem1}, which separate
the singlet-state quantum correlations
from all LHV correlations for $\theta \in
\bigl(0,\frac{\pi}{3}\bigr)$.

We have also explored hypotheses that would refine and unify these
results further: the weak and strong hemispherical colouring
maximality hypotheses.  These state that the LHV defined by the
simplest spherical colouring, with opposite hemispheres coloured
oppositely, maximizes the LHV anticorrelations for a continuous range
of $\theta > 0$, either among LHVs with perfect anticorrelation at $\theta=0$ (the
weak case) or without any restriction (the strong case).

We should note here that the intuition supporting the WHCMH relates
specifically
to colourings in two or more dimensions, where there seems no
obvious way of constructing colourings that vary over small
scales in a way that is regular enough to produce very strong
(anti) correlations for small $\theta$.

On the other hand, the one-dimensional analog of the WHCMH -- that the
strongest
anticorrelations for colourings on the circle arise from
colouring opposite half-circles oppositely -- is easily seen to be false.
For $n$ odd, the colouring $a (\epsilon ) =-b(\epsilon)= (-1)^{\lfloor \frac{ n
    \epsilon }{  \pi} \rfloor } $ with $\epsilon\in[0,2\pi]$ is antipodal and
is perfectly anticorrelated
for $\theta = \frac{2 \pi }{n}$.

Although it underlines that the hemispherical colouring hypotheses
are non-trivial, this distinction between one and higher dimensions is consistent
with what is known about other colouring problems in geometric
combinatorics \cite{bukh,de2008fourier}.
The intuition that colouring $1$ should be optimal, because it
solves the isoperimetric problem of finding the coloured region
with half the area of the sphere that has the shortest boundary, 
remains suggestive.  Verifying the WHCMH and the SHCMH look
at first sight like simple classical problems in geometry and combinatorics
that can be stated quite independently of quantum theory. They have
many interesting generalisations \footnote{For example, among
non-antipodal bipartite colourings of the sphere in which the black
region has area $A< 2 \pi$, which colouring(s) produce maximal
correlation?    Or, consider a general region $R$ of volume $V$
in $\mathbb{R}^n$, and define $p_{\epsilon} (R)$ to be the
probability that, given a randomly chosen point $x \in R$,
and a randomly chosen point $y$ such that $d(x,y) = \epsilon$,
we find that $y \in R$.   Do the
balls maximize this probability, for any
given sufficiently small $\epsilon$?}. Nonetheless, as far as we are aware,
these questions have not been seriously studied by
pure mathematicians to date, although some intriguing relatively recent
results \cite{bukh,de2008fourier} on colourings
in $\mathbb{R}^n$ encourage hope that proof methods
could indeed be found.
We thus simply state the WHCMH and the SHCMH as interesting and
seemingly plausible hypotheses to be investigated
further rather than offering them as conjectures,
preferring to reserve the latter terms for propositions for which
very compelling evidence has been amassed.

We would like to stress what we see as a key insight
deserving further exploration, namely that stronger
and more general Bell inequalities 
could in principle be proven by results about continuous
colourings, rather than restricting to colourings of discrete
sets.    While we have focussed on the simplest case of
projective measurements of pairs of qubits, this observation
of course applies far more generally.
We hope our work will stimulate further investigation
of the WHCMH and the SHCMH and related colouring
problems, which seem very interesting in their own
right, and in developing further
this intriguing link between
pretty and natural questions in geometric combinatorics and
measures of quantum nonlocality.

We have considered here the ideal case in which Alice and Bob
share a maximally entangled pure state and are able to carry
out perfect projective measurements about axes specified
with perfect precision.   For a range of 
non-zero $\theta$, our results 
show a finite separation between the predictions of quantum
theory and LHVTs.   As is the case for CHSH and other Bell tests,
they can thus also be applied (within a certain parameter
range) to realistic experiments in which the entangled 
state is mixed and measurements can only be 
approximately specified.   In particular, they offer 
new methods for exploring the range of parameters for
which the correlations defined by 
rotationally symmetric Werner states can be 
distinguished from those of any LHVT \cite{W89,AGT06,V08,BCPSW14}.   It would
be interesting to explore this further.   

Finally, but importantly, we would like to note earlier work on
related questions. In a pioneering paper, \.{Z}ukowski \cite{zukowskipla} considered generalised
Bell and GHZ tests for maximally entangled quantum states that
involve all possible axis choices, and gave an elegant proof that the quantum
correlations can be distinguished from all possible LHVT
correlations by a weighted average measure of correlation
functions.  For the bipartite case, our work investigates
the gap between quantum and LHVT correlations at
each axis angle separation.  This allows one to define infinitely many
generalised Bell tests corresponding to different
weighted averages of correlation functions.  It would be 
interesting to characterise the space of all such tests and
its boundaries.    

References \cite{liang,shadbolt,WB12} investigate \emph{inter alia}
Bell-CHSH experiments in which the axes are initially
chosen randomly, and the same axes are used 
repeatedly throughout a given experimental run. 
Reference \cite{liang} shows that such experiments lead 
to Bell inequality violations a significant fraction of 
the time when pairs of random local measurements are chosen.
References \cite{shadbolt,WB12} show that by considering triads 
of random local measurements, constrained to be  mutually unbiased,
for which Alice's axes are not perfectly aligned to Bob's axes,
the violation of a CHSH inequality is guaranteed on a two 
qubit maximally entangled state.
Their scenarios are significantly different from ours. 
In our scenario, the axes are chosen randomly and independently
for each measurement, and (in the ideal case) Alice and Bob have
the ability to define their axis choices precisely with respect
to the same reference frame.    
The goals are also different: 
References \cite{liang, shadbolt,WB12} show that Bell inequality violation
can be demonstrated even when Alice and Bob do not have a 
shared reference frame; our aim is to establish
new Bell inequalities rather than to exploit the power of 
known inequalities.   It would be interesting to
explore possible connections, nonetheless. 

After completing this work, our attention was also drawn to 
a related question considered in \cite{AMRSV13}; see 
Appendix \ref{appendix questions} for discussion.  

\begin{acknowledgments}
We thank Boris Bukh for very helpful discussions and for drawing our
attention to Refs. \cite{bukh,de2008fourier}.  A.K. was partially
supported by a grant from the John Templeton Foundation and by
Perimeter Institute for Theoretical Physics. Research at Perimeter
Institute is supported by the Government of Canada through Industry
Canada and by the Province of Ontario through the Ministry of
Research and Innovation.
D.P.-G. thanks Tony Short, Boris Groisman and Jonathan Barrett for helpful
discussions, and acknowledges financial support from CONACYT M\'exico and
partial support from Gobierno de Veracruz.

\end{acknowledgments}

\appendix
\section{\label{appendix proofs} Proofs of the theorem and lemmas}

\subsection{Proof of Lemma \ref{lemma1}}
From the CHSH inequality,
\begin{equation}
\label{eq:5}
\bigl\lvert C(0,0)+C(1,1)+C(1,0)-C(0,1)\bigr\rvert \leq 2 \, ,
\end{equation}
in the case in
which the measurements $A=0$, $A=1$ and $B=0$ correspond to
projections on states with Bloch vectors separated from each other by
the same angle $\theta\in\bigl(0,\frac{2\pi}{3}\bigr]$, Bob's
measurement $B=1$ is the same as Alice's measurement $A=0$ and the
outcomes are described by LHVTs satisfying (\ref{eq:14}), we obtain
after averaging over random rotations of the Bloch sphere that
$\lvert 3C_x(\theta)-C_x(0)\rvert\leq 2$. Then, the result follows because, as shown in the main text, we have $C_x(0)=-1+2\gamma$.\qed

\subsection{Proof of Lemma \ref{lemma2}}
\begin{figure}
\includegraphics{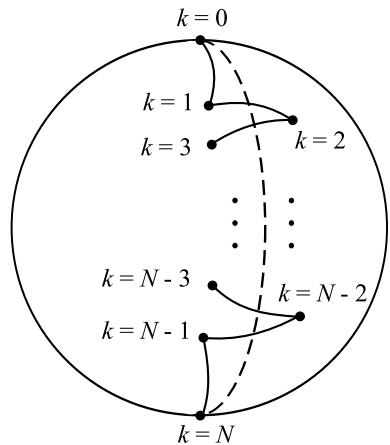}
 \caption{\label{figzigzagsm}Diagram of the measurements performed by Alice and
Bob that are used in the proof of Lemma~\ref{lemma2}. Alice's and Bob's
measurements $k$ are the same, for $k=0,1,\ldots,N-1$ and $N\geq 2$; these are
projections onto the states $\lvert\xi_k\rangle$ and correspond to points in
the Bloch sphere with label $k$. These points form a zigzag path crossing the
dashed great circle. The state $\lvert\xi_N\rangle$ is antipodal to
$\lvert\xi_0\rangle$ and represents the measurement $k=0$ with reversed
outcomes. The solid lines represent arcs of great circles with the same angle
$\theta>\frac{\pi}{N}$ that connect adjacent points.  If
$\theta=\frac{\pi}{N}$, all these points are on the same great circle.}
\end{figure}

From the Braunstein-Caves inequality, Eq.~(\ref{eq:7}), we have that
\begin{multline}
\label{eq:18}
I_N = \biggl\lvert \sum_{k=0}^{N-1} C(k,k)  + \sum_{k=0}^{N-1} C(k+1,k)\biggr\rvert \leq
2N-2
\, , \end{multline} with the convention that measurement choice $N$ is
measurement choice $0$ with reversed outcomes. We consider the case in which
Alice's and Bob's measurement $k$ are the same, for $k=0,1,\ldots,N-1$  and
$N\geq 2$, and their
outcomes are described by LHVTs satisfying (\ref{eq:14}) and (\ref{eq:17.1}),
which then also satisfy $C_x(0)=-1+2\gamma$. If we take measurement
$k$ to be of the projection onto the state $\lvert\xi_k\rangle$ so
that the states $\lbrace\lvert\xi_{k}\rangle\rbrace_{k=0}^{N-1}$ are
along a great circle on the Bloch sphere with a separation angle
$\theta=\frac{ \pi }{ N} $ between $\lvert\xi_k\rangle$ and
$\lvert\xi_{k+1}\rangle$ for $k=0,1,\ldots,N-2$, for example
$\lvert\xi_k\rangle=\cos \bigl(\frac{k\pi}{2N}\bigr) \lvert 0\rangle +
\sin \bigl(\frac{k\pi}{2N}\bigr) \lvert 1\rangle$, and average over
random rotations of the Bloch sphere, this gives 
\begin{equation}
\label{eq:18.1}
\bigl\lvert  NC_x(0)  + N C_x(\theta)\bigr\rvert \leq
2N-2
\, . \end{equation}
Since $C_x(0)=-1+2\gamma$, it follows that $ C_x \bigl(
\frac{\pi}{N} \bigr) \geq -1 + \frac{2}{N} -2\gamma=C_1\bigl( \frac{\pi}{N}
\bigr)-2\gamma\, . $ Similarly, if we
take the states $\lbrace\lvert\xi_{k}\rangle\rbrace_{k=0}^{N-1}$ to be
along a zigzag path crossing a great circle on the Bloch sphere with
a separation angle $\theta>\frac{\pi}{N}$ between $\lvert\xi_k\rangle$ and
$\lvert\xi_{k+1}\rangle$ for $k=0,1,\ldots,N-2$, in such a way that
the angle separation between $\lvert\xi_{N-1}\rangle$ and the state
with Bloch vector antiparallel to that one of $\lvert\xi_0\rangle$ is
also $\theta$ (see Fig.~\ref{figzigzagsm}), we obtain after averaging over
random rotations of the
Bloch sphere that $ C_x ( \theta ) \geq -1 + \frac{2}{N}-2\gamma=C_1\bigl(
\frac{\pi}{N} \bigr)-2\gamma$.\qed

\subsection{Proof of Theorem \ref{theorem1}}
\begin{figure}
\includegraphics{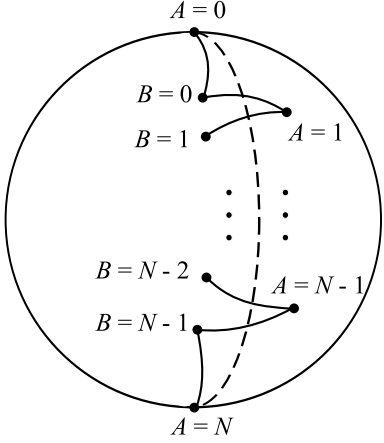}
 \caption{\label{figzigzagdm}Diagram of the measurements performed by Alice and
Bob that are used in the proof of Theorem \ref{theorem1}. Alice's and Bob's measurements
$A$ and $B$ are projections onto the state $\lvert\xi_A\rangle$ and
$\lvert\chi_B\rangle$ and correspond to points in the Bloch sphere with labels
$A$ and $B$, respectively, for $A,B\in\lbrace 0,1,\ldots,N-1\rbrace$ and $N\geq
2$. These points form a zigzag path crossing the dashed great circle. The state
$\lvert\xi_N\rangle$ is antipodal to $\lvert\xi_0\rangle$ and represents
Alice's measurement $A=0$ with reversed outcomes. The solid lines represent
arcs of great circles with the same angle $\theta>\frac{\pi}{2N}$ that connect
adjacent points.  If $\theta=\frac{\pi}{2N}$, all these points are on the same
great circle.}
\end{figure}

Consider the Braunstein-Caves inequality, Eq.~(\ref{eq:7}), in the
case in which Alice's and Bob's measurement outcomes are described by LHVTs
satisfying (\ref{eq:14}).
Let Alice's and Bob's measurements $k$ correspond to the projections onto
the states $\lvert\xi_k\rangle$
and $\lvert\chi_k\rangle$, respectively, for $k=0,1,\ldots,N-1$ and $N\geq 2$.
Let the angle along the great circle in the Bloch sphere passing through the
states
$\lvert\xi_k\rangle$ and $\lvert\chi_{k}\rangle$  be $\theta$, for
$k=0,1,\ldots,N-1$. Similarly, let the angle along the great circle passing
through $\lvert\chi_{k}\rangle$ and $\lvert\xi_{k+1}\rangle$ be $\theta$ for
$k=0,1,\ldots,N-1$, with the convention that the state $\lvert\xi_{N}\rangle$
has Bloch vector antiparallel to that one of $\lvert\xi_0\rangle$. If
$\theta=\frac{\pi}{2N}$, all these states are on the same great circle
beginning at $\lvert\xi_0\rangle$ and ending at $\lvert\xi_N\rangle$. If
$\theta>\frac{\pi}{2N}$, the states can be accommodated on a zigzag path
crossing the great circle that goes from $\lvert\xi_0\rangle$ to
$\lvert\xi_N\rangle$ (see Fig.~\ref{figzigzagdm}). Thus, from the
Braunstein-Caves inequality, after averaging over random rotations of the Bloch sphere, we have
$C_1\bigl(\frac{\pi}{2N}\bigr)=-1+\frac{1}{N}\leq C_x(\theta)  \leq
1-\frac{1}{N}=- C_1\bigl(\frac{\pi}{2N}\bigr)$, for
$\theta\geq\frac{\pi}{2N}$.\qed

\subsection{Proof of Lemma \ref{lemma3}}
Consider a colouring $x\in\mathcal{X}$ and an angle
$\theta\in\bigl(\frac{\pi}{M+1},\frac{\pi}{M}\bigr] $ for an integer $M\geq 2$
such that $C_x ( \theta) < C_1 (\theta)$ or $C_x ( \theta) > -C_1 (\theta)$.
From Theorem \ref{theorem1} and the fact that
$C_x\bigl(\frac{\pi}{2}\bigr)=C_1\bigl(\frac{\pi}{2}\bigr)=0$, it must be that
$\theta\neq\frac{\pi}{M}$ if $M$ is even. We define the angles $\theta_j\equiv
\frac{\pi}{M+1-j}-\theta$ with $j=1,2,\ldots,M-1$.
Considering the cases $M$ even and $M$ odd, and using that
$\theta\neq\frac{\pi}{M}$ if $M$ is even,  it is straightforward to obtain that
$0\leq\theta_j<\theta$ if $j<\frac{M}{2}+1$ and $\frac{\pi}{2}>\theta_j>\theta$
if $j\geq \frac{M}{2}+1$. Now consider the Braunstein-Caves inequality, Eq.~(\ref{eq:7}), in the
case in which
Alice's and Bob's measurement outcomes are described by LHVTs satisfying (\ref{eq:14}).
Let Alice's and Bob's measurements $k$ correspond to the projections onto
the
states $\lvert\xi_k\rangle$
and $\lvert\chi_k\rangle$, respectively, for $k=0,1,\ldots,N-1$ and $N\equiv
M+1-j$. Since $1\leq j\leq M-1$, we have $2\leq N\leq M$. Let all these states
be on the great circle in the Bloch sphere that passes through the states
$\lvert\xi_0\rangle$ and $\lvert\xi_N\rangle$, with the convention that the
state $\lvert\xi_{N}\rangle$ has Bloch vector antiparallel to that one of
$\lvert\xi_0\rangle$. Let the angles between $\lvert\xi_k\rangle$ and
$\lvert\chi_k\rangle$, and between $\lvert\chi_k\rangle$ and
$\lvert\xi_{k+1}\rangle$ along this great circle be $\theta$ and $\theta_{j}$,
respectively. For example,
$\lvert\xi_k\rangle=\cos \bigl(\frac{k\pi}{2N}\bigr) \lvert 0\rangle +
\sin \bigl(\frac{k\pi}{2N}\bigr) \lvert 1\rangle$ and $\lvert\chi_k\rangle=\cos
\bigl(\frac{k\pi}{2N}+\frac{\theta}{2}\bigr) \lvert 0\rangle +
\sin \bigl(\frac{k\pi}{2N}+\frac{\theta}{2}\bigr) \lvert 1\rangle$, for
$k=0,1,\ldots,N-1$. From the Braunstein-Caves inequality, after
averaging over random rotations of the Bloch sphere, we obtain
$-1+\frac{1}{N}\leq\frac{1}{2}\bigl(C_x(\theta)+C_x(\theta_{j})\bigr)\leq
1-\frac{1}{N}$. Since the average angle
$\bar{\theta}_{j}\equiv\frac{1}{2}(\theta+\theta_{j})$ satisfies
$\bar{\theta}_{j}=\frac{\pi}{2(M+1-j)}=\frac{\pi}{2N}$ and
$C_1(\frac{\pi}{2N})=-1+\frac{1}{N}$, we have
$C_1\bigl(\bar{\theta}_{j}\bigr)\leq\frac{1}{2}\bigl(C_x(\theta)+C_x(\theta_{j})\bigr)\leq
-C_1\bigl(\bar{\theta}_{j}\bigr)$. Since $C_1(\theta)$ is a linear function of
$\theta$, it follows that $C_x(\theta_{j})>C_1(\theta_{j})$ if
$C_x(\theta)<C_1(\theta)$. Similarly, $C_x(\theta_{j})<-C_1(\theta_{j})$ if
$C_x(\theta)>-C_1(\theta)$.\qed

\subsection{Proof of Lemma \ref{lemma4}}
Let $x\in\mathcal{X}$ be any colouring and
$\theta\in\bigl(0,\frac{\pi}{3}\bigr)$. We first consider the case
$\theta\in\bigl[\frac{\pi}{4},\frac{\pi}{3}\bigr)$. From Theorem \ref{theorem1}, we have $C_1\bigl(\frac{\pi}{4})\leq C_x(\theta)\leq
-C_1\bigl(\frac{\pi}{4})$. The quantum correlation for the singlet state is
$Q(\theta)=-\cos{\theta}$. Since $Q(\theta)$ is a strictly increasing function
of $\theta$, we have
$Q(\theta)<Q\bigl(\frac{\pi}{3}\bigr)=-\frac{1}{2}=C_1\bigl(\frac{\pi}{4}\bigr)$
for $\theta<\frac{\pi}{3}$. Therefore, $Q(\theta)<C_x(\theta)<-Q(\theta)$ for
$\theta\in\bigl[\frac{\pi}{4},\frac{\pi}{3}\bigr)$. Similarly, it is easy to
see that $Q(\theta)<C_x(\theta)<-Q(\theta)$ for
$\theta\in\bigl[\frac{\pi}{6},\frac{\pi}{4}\bigr)$.
Now we consider the case $\theta\in\bigl(0,\frac{\pi}{6}\bigr)$. We define
$N=\lceil \frac{\pi}{2\theta}\rceil$. It follows that
$\theta\in\bigl[\frac{\pi}{2N},\frac{\pi}{2(N-1)}\bigr)$ for an integer $N\geq
4$.
From Theorem \ref{theorem1}, we have $-1+\frac{1}{N}=
C_1\bigl(\frac{\pi}{2N}\bigr)\leq C_x(\theta)\leq
-C_1\bigl(\frac{\pi}{2N}\bigr)=1-\frac{1}{N}$. From the Taylor series
$Q(\theta)=-1+\frac{\theta^2}{2}-\frac{\theta^4}{4!}+\frac{\theta^6}{6!}-\cdots$,
it is easy to see that $Q(\theta)<-1+\frac{\theta^2}{2}$ for
$0<\theta<\sqrt{30}$. Thus, we have
$Q\bigl(\frac{\pi}{2(N-1)}\bigr)<-1+\frac{1}{2}\bigl(\frac{\pi}{2(N-1)}\bigr)^2$.
Since $N^2>\bigl(\frac{\pi^2}{8}+2\bigr)N-1$, it follows that
$(N-1)^2>\frac{\pi^2}{8}N$, which implies that
$-1+\frac{1}{N}>-1+\frac{1}{2}\bigl(\frac{\pi}{2(N-1)}\bigr)^2$. It follows
that $C_x(\theta)> Q\bigl(\frac{\pi}{2(N-1)}\bigr)$. Since $Q(\theta)$ is a
strictly increasing function of $\theta$ and $\theta<\frac{\pi}{2(N-1)}$, we
have $Q\bigl(\frac{\pi}{2(N-1)}\bigr)>Q(\theta)$. Thus, we have
$C_x(\theta)>Q(\theta)$. Similarly, we have $C_x(\theta)<-Q(\theta)$.\qed

\section{Related questions for exploration}
\label{appendix questions}

As mentioned in the main text, some interesting related questions involving non-local games with continuous inputs
have been considered in \cite{AMRSV13}.   In particular, in the third
game considered in \cite{AMRSV13},  Alice and Bob are given uniformly
distributed Bloch sphere vectors, $\vec{r}_A$ and $\vec{r}_B$, and aim to maximise the
probability of producing outputs that are anticorrelated if 
$\vec{r}_A \cdot \vec{r}_B \geq 0$ or correlated 
if $\vec{r}_A \cdot \vec{r}_B < 0$.   Aharon \emph{et al}. suggest that the 
LHV strategy defined by opposite hemispherical colourings is optimal,
though they give no argument.    They also suggest that the quantum
strategy given by sharing a singlet and carrying out measurements
corresponding to the input vectors is optimal, based on evidence
from semi-definite programming.   Equation (\ref{eq:new}) shows
that this is the case for all $\theta$, and so in particular
for the average advantage in the game considered, if Alice
and Bob are restricted to outputs defined by projective measurements
on a shared pair of qubits.   Our earlier results also prove
that there is a quantum advantage for all $\theta$ in the
range $0 < \theta < \frac{\pi}{3}$, 
and hence for many versions of this game defined by
a variety of probability distributions for the inputs.  

We show Eq.~(\ref{eq:new}) below. First, we compute the average outcome probabilities when Alice and Bob apply local projective measurements on a two-qubit state $\rho$, for measurement bases defined by Bloch vectors separated by an angle $\theta$. The average is taken over random rotations of these vectors in the Bloch sphere, subject to the angle separation $\theta$. Then, we compute the quantum correlations.

Consider a fixed pair of pure qubit states $\lvert 0\rangle$ and $\lvert\chi\rangle=\cos\bigl(\frac{\theta}{2}\bigr)\lvert 0\rangle+\sin\bigl(\frac{\theta}{2}\bigr)\lvert 1\rangle$ for Alice's and Bob's measurements, respectively, corresponding to outcomes `+1'. A general state for Bob's measurement separated by an angle $\theta$ with respect to a fixed state $\lvert 0\rangle$ for Alice's measurement is obtained by applying the unitary $R_z(\omega)$ that corresponds to a rotation of an angle $\omega\in[0,2\pi]$ around the $z$ axis in the Bloch sphere, which only adds a phase to the state $\lvert 0\rangle$. Then, after applying $R_z(\omega)$, a general pure product state $\lvert\xi_{\vec{a}}\rangle\otimes\lvert\chi_{\vec{b}}\rangle$ of two qubits with Bloch vectors separated by an angle $\theta$ is obtained by applying the unitary $R_z(\phi)R_y(\epsilon)$ that rotates the Bloch sphere around the $y$ axis by an angle $\epsilon\in[0,\pi]$ and then around the $z$ axis by an angle $\phi\in[0,2\pi]$. Thus, we have $\lvert\xi_{\vec{a}}\rangle\otimes\lvert\chi_{\vec{b}}\rangle=U_{\phi,\epsilon,\omega}\lvert 0 \rangle \otimes U_{\phi,\epsilon,\omega}\lvert\chi\rangle$, with $U_{\phi,\epsilon,\omega}=R_z(\phi)R_y(\epsilon) R_z(\omega)$. This is a general unitary acting on a qubit, up to a global phase. Therefore, we can parametrize this unitary by the Haar measure $\mu$ on $\text{SU(2)}$, hence, we have $\lvert\xi_{\vec{a}}\rangle\otimes\lvert\chi_{\vec{b}}\rangle=U_{\mu}\lvert 0\rangle\otimes U_{\mu}\lvert \chi\rangle$.

After taking the average, the probability that both Alice and Bob obtain the outcome `+1' is
\begin{eqnarray}
\label{eq:quantumcorrelation}
\!P(\!+\!\!+\!\!\vert\theta)&\!\!=\!\!&\int\!\!\! d\mu\text{Tr}\Bigl(\rho\bigl(\lvert \xi_{\vec{a}}\rangle\langle \xi_{\vec{a}}\rvert\otimes\lvert \chi_{\vec{b}}\rangle\langle\chi_{\vec{b}}\rvert\bigr)\Bigr)\nonumber\\
&\!\!=\!\!&\int\!\!\! d\mu\text{Tr}\Bigl(\rho \bigl(U_{\mu}\otimes U_{\mu}\bigr)\bigl(\lvert 0\rangle\langle 0\rvert\otimes \lvert \chi\rangle\langle\chi\rvert \bigr)\bigl(U_{\mu}^{\dagger}\otimes U_{\mu}^{\dagger}\bigr)\!\Bigr)\nonumber\\
&\!\!=\!\!&\text{Tr}\biggl(\!\int\!\!\! d\mu\bigl(U_{\mu}^{\dagger}\otimes U_{\mu}^{\dagger}\bigr)\rho\bigl(U_{\mu}\otimes U_{\mu}\bigr)\bigl(\lvert 0\rangle\langle 0\rvert\otimes \lvert \chi\rangle\langle\chi\rvert \bigr)\!\!\biggr)\nonumber\\
&\!\!=\!\!&\text{Tr}\Bigl(\tilde{\rho}\bigl(\lvert 0\rangle\langle 0\rvert\otimes \lvert \chi\rangle\langle\chi\rvert \bigr)\Bigr),
\end{eqnarray}
where in the third line we used the linearity and the cyclicity of the trace and in the fourth line we used the definition $\tilde{\rho}\equiv\int d\mu\bigl(U_{\mu}^{\dagger}\otimes U_{\mu}^{\dagger}\bigr)\rho\bigl(U_{\mu}\otimes U_{\mu}\bigr)$. The state $\tilde{\rho}$ is invariant under a unitary transformation $U\otimes U$, for any $U\in\text{SU}(2)$. The only states with this symmetry are the Werner states \cite{W89}, which for the two-qubit case have the general form
\begin{equation}
\label{eq:wernerstate}
\tilde{\rho}\!=\!r \lvert\Psi^-\rangle\langle\Psi^-\rvert+\frac{1-r}{3}\bigl(\lvert\Psi^+\rangle\langle\Psi^+\rvert+\lvert\Phi^+\rangle\langle\Phi^+\rvert+\lvert\Phi^-\rangle\langle\Phi^-\rvert\bigr)\!,
\end{equation}
with $0\leq r\leq 1$. Thus, from Eqs.~(\ref{eq:quantumcorrelation}) and (\ref{eq:wernerstate}), we obtain
\begin{equation}
\label{eq:pp}
P(++\vert\theta)=\frac{1-r}{3}+\frac{4r-1}{6}\sin^2\Bigl(\frac{{\theta}}{2}\Bigr).    
\end{equation}

Since the projectors corresponding to Alice and Bob obtaining outcomes `-1' are obtained by a unitary transformation of the form $U\otimes U$ on $\lvert 0\rangle\otimes\lvert\chi\rangle$,  with $U\in\text{SU}(2)$, then from Eq.~(\ref{eq:quantumcorrelation}) we see that after integrating over the Haar measure on SU(2), we obtain $P(--\vert\theta)=P(++\vert\theta)$. 

Thus, the average quantum correlation is $Q_{\rho}(\theta)=4P(++\vert\theta)-1$, which from Eq.~(\ref{eq:pp}) gives
\begin{equation}
\label{eq:qc}
Q_{\rho}(\theta)=-\Bigl(\frac{4r-1}{3}\Bigr)\cos{\theta}.
\end{equation}
Then, Eq.~(\ref{eq:new}) follows because $0\leq r\leq 1$.

\section{Numerical results}
\label{appendix nr}
We investigated the WHCMH numerically by computing
the correlation $C_x(\theta)$ for various colouring
functions that satisfy the antipodal property (\ref{eq:antipodalx}), the
condition
(\ref{eq:17}), and
that have azimuthal symmetry.  These colourings are illustrated in
Fig.~\ref{fig2} and defined in Appendix~\ref{appendix A}.

\begin{figure}
\includegraphics{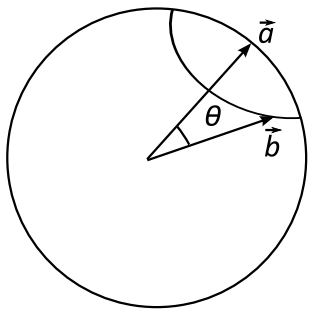}
\caption{\label{fig1}Alice's and Bob's measurement axes $\vec{a}$ and
  $\vec{b}$ form an angle $\theta$. The spherical coordinates of
  $\vec{a}$ and $\vec{b}$ are $(\epsilon,\phi)$ and $(\alpha,\beta)$,
  respectively, related by Eqs.~(\ref{eq:10}) and (\ref{eq:11}).
  Equation~(\ref{eq:14}) computes the correlation $C_x(\theta)$
  by (i) integrating the colouring
  function $b_x\bigl(\vec{b}\!~\bigr)$ over the circle on the
  sphere generated by $\vec{b}$ (parametrized by the angle $\omega$
  in Eqs.~(\ref{eq:10}) and (\ref{eq:11})) and (ii) integrating the
   colouring function $a_x(\vec{a})$ over the sphere
  generated by $\vec{a}$. A general correlation $C(\theta)=\int_{\mathcal{X}}dx\mu(x)C_x(\theta)$ is computed by integrating over the probability
  distribution $\mu(x)$ of the colourings satisfying the antipodal property (\ref{eq:antipodalx}).}
\end{figure}

We define $(\epsilon,\phi)$ as the spherical coordinates 
of
$\vec{a}$ and $(\alpha,\beta)$ as those of $\vec{b}$; where
$\epsilon,\alpha\in[0,\pi]$ are angles from the north pole and
$\phi,\beta\in[0,2\pi]$ are azimuthal angles. The vectors $\vec{a}$ and $\vec{b}$
are separated by a fixed angle $\theta$. The set of possible values of $\vec{b}$ around the fixed axis $\vec{a}$ generate a circle parametrized by an angle $\omega$ (see Fig.~\ref{fig1}). The spherical
coordinates $(\alpha,\beta)$ for a point $\vec{b}$ with angular
coordinate $\omega$ on this circle are:
\begin{eqnarray}
\label{eq:10}
\alpha &&= \arccos(\cos\theta\cos\epsilon -
\sin\theta\sin\epsilon\cos\omega),\\
\label{eq:11}
\beta
&&=\!\biggl[\!\phi+k_\omega\arccos\Bigl({\frac{\cos\epsilon\sin\theta\cos\omega+\sin\epsilon\cos\theta}{\sin\alpha}}\Bigr)\!\biggr]\text{\!\!
mod\! } 2\pi,\nonumber\\
\end{eqnarray}
where $k_\omega=1$ if $0\leq\omega\leq\pi$ and $k_\omega=-1$ if $\pi<\omega\leq
2\pi$. Notice that $\beta$ is undefined for $\alpha\in\lbrace 0,\pi\rbrace$.

Equations~(\ref{eq:10}) and
(\ref{eq:11}) were used to compute the double integral in
(\ref{eq:14}). The integral with respect to the angle $\omega$ was
performed analytically. Thus, the correlations $C_x(\theta)$ were
reduced to a sum of terms that include single integrals with respect
to the polar angle $\epsilon$; the obtained expressions are given in
Appendix~\ref{appendix B}. The single integrals with respect to
$\epsilon$ were computed numerically with a program using the software
\emph{Mathematica}, which we provide as supplemental material.

\begin{figure}
\includegraphics{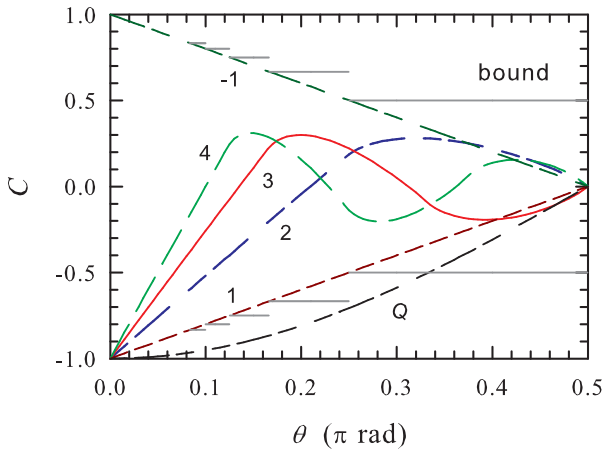}
\caption{\label{fig3} Correlations computed with (\ref{eq:14}), subject to the constraint (\ref{eq:17}), for the colouring
functions $a_x$ shown
  schematically in Fig.~\ref{fig2} and defined in
  Appendix~\ref{appendix A}. The correlations for colouring 2, 3 and 4
  are blue dot-dashed, red solid and green dashed curves,
  respectively. The black dot-dash-dotted curve represents the singlet-state
quantum correlation $ Q ( \theta )$. The  dark red
  dotted and dark green dash-dotted curves show respectively the colouring $1$
correlation, $ C_1 (\theta )$, and anticorrelation, $ -C_1 (\theta )$. The
gray solid straight lines show the bounds given by Theorem \ref{theorem1}, for $\theta\geq
\frac{\pi}{12}$.}
\end{figure}

Our results are plotted in Fig.~\ref{fig3}; they are consistent with
the WHCMH.  They also show that $\theta_{\rm max}^{\text{w}} < \frac{\pi}{2} $,
because
they show that there exists a colouring $x$ with $C_x(\theta)<C_1(\theta)$ for
some
angles
$\theta\in\bigl(0,\frac{\pi}{2}\bigr)$, namely
colouring 3 for
angles $\theta\in\bigl[0.405\pi,\frac{\pi}{2}\bigr)$.

\begin{figure}
\includegraphics{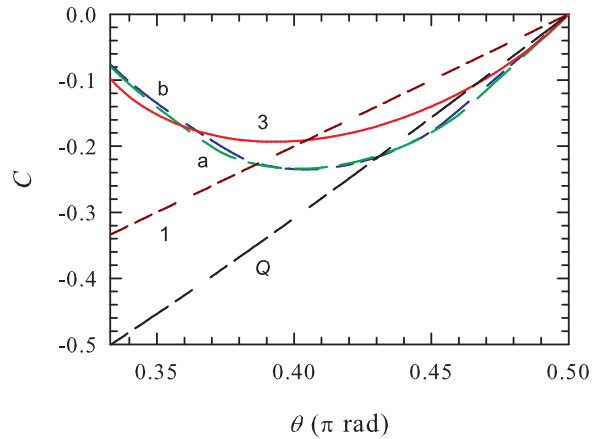}
 \caption{\label{fig4} Correlations obtained for colouring
$3_\delta$, defined in Appendix~\ref{appendix A}, for $\delta=-0.038\pi$ (a,
green
dashed curve) and $\delta=-0.046\pi$ (b, blue
dot-dashed curve); for colourings
3, 1 and the singlet-state quantum correlation $Q(\theta)$ (red solid, dark red
dotted and black
dot-dash-dotted curves, respectively).}
\end{figure}

Another interesting result is that there exist colourings that produce
correlations $ C_x(\theta)< Q(\theta)$ for
$\theta$ close to $\frac{\pi}{2}$: colouring 3 for angles
$\theta\in\bigl[0.467\pi,\frac{\pi}{2}\bigr)$. It is interesting to find other
colourings whose correlations satisfy $ C_x(\theta)<C_1(\theta)$ and $
C_x(\theta)<Q(\theta)$ for angles $\theta$ closer to zero. For this purpose,
we consider colouring $3_\delta$, which is defined in
Appendix~\ref{appendix A} and consists of a small variation of
colouring 3 in terms of the parameter $\delta$. Colouring $3_\delta$
reduces to colouring $3$ if $\delta=0$. For values of $\delta$ in the
range $\bigl[-\frac{\pi}{18},\frac{\pi}{24}\bigr]$, we obtained that the
smallest angle $\theta$
for which $ C_{3_\delta}(\theta)<C_1(\theta)$ is achieved for $\delta=-0.038
\pi$, in which case
we have that $ C_{3_{-0.038 \pi}}(\theta)<C_1(\theta)$ for
$\theta\in\bigl[0.386 \pi,\frac{\pi}{2}\bigr)$. We also obtained that the
smallest angle $\theta$ for which $ C_{3_\delta}(\theta)< Q(\theta)$ is
achieved for $\delta=-0.046\pi$, in which case we have that
$C_{3_{-0.046\pi}}(\theta)< Q(\theta)$ for
$\theta\in[0.431\pi,\frac{\pi}{2}\bigr)$ (see
Fig.~\ref{fig4}).

Our numerical results imply the bound $\theta_{\text{max}}^{\text{w}}\leq 0.386
\pi$.
They also imply that $\theta_{\text{max}}^{\text{s}}\leq 0.375\pi$, because
$C_2(\theta)> -C_1(\theta)$
for $\theta\in\bigl(0.375\pi,\frac{\pi}{2}\bigr)$, and $C_1(\theta)\leq
C_x(\theta)\leq -C_1(\theta)$
for $x=2,3,4,3_\delta$ and $\theta\in[0,0.375\pi]$. 

The slightly improved bound $\theta_{\text{max}}^{\text{s}}\leq 0.345\pi$ was obtained in \cite{DPGthesis} from a variation of colouring $2$, colouring $2_{\Delta}$, in which the polar angle defining the boundary between the black and white regions in the northern hemisphere (see Fig.~\ref{fig2}) is reduced by the angle $\Delta\in\bigl[0,\frac{\pi}{12}\bigr]$.

In order to confirm analytically the numerical observation that there exist
colouring functions $x\in\mathcal{X}$ such that $ C_x(\theta)<
Q(\theta)$ for $\theta$ close to $\frac{\pi}{2}$, we computed
$C_3\bigl(\frac{\pi}{2}-\tau\bigr)$ for $0\leq\tau\ll 1$ to order
$\mathcal{O}(\tau^2)$. The computation is presented in Appendix~\ref{appendix C}.   We obtain
\begin{equation}
\label{eq:22}
C_3\Bigl(\frac{\pi}{2}-\tau\Bigr)=-1.5\tau+\mathcal{O}(\tau^2).
\end{equation}
On the other hand, the quantum correlation gives
$Q\bigl(\frac{\pi}{2}-\tau\bigr)=-\cos\bigl(\frac{\pi}{2}-\tau\bigr)=-\tau+\mathcal{O}(\tau^3).$
Thus, we see that for $\tau$ small enough, indeed
$ C_3\bigl(\frac{\pi}{2}-\tau\bigr)<Q\bigl(\frac{\pi}{2}-\tau\bigr)$.

Further numerical investigations of the WHCMH and SHCMH
might well shed further light on the questions we explore here.   For example,
one could define
an antipodal colouring function $x$ as the sign of a sum
of spherical harmonics,
${\rm sgn} ( \sum_{m=-l}^l \sum_{l=0}^L  a_{lm} Y_{lm} (\epsilon, \phi ))$,
where the coefficients $a_{lm}$ are variable parameters,
and then search for the minimum value of $C_x ( \theta )$,
for any given $\theta$, among such functions by optimizing with respect to the
$a_{lm}$.
As an ansatz, one might assume that components corresponding to
spherical harmonics that oscillate
rapidly compared to $\theta$ are relatively negligible,
given that the colourings defined by such functions contain black and
white areas small compared to $\theta$ everywhere on the sphere, giving a
contribution to the correlation
very close to zero.   This would allow searches over a finite set of
parameters, for any given $\theta$, while the ansatz itself can be
tested by finding how the maximum changes with increasing $L$.

\subsection{\label{appendix A} Definitions of the colouring functions}

In general, a colouring function $a_x$ with azimuthal symmetry can be
defined in terms of the set $\mathcal{E}_x$ in which it takes the
value 1 as follows:
\begin{eqnarray}
\label{equationC4}
a_x(\epsilon)\equiv\left\{\begin{array}{rl}
1 & \text{if } \epsilon\in \mathcal{E}_x,\nonumber\\
-1 & \text{if } \epsilon\in[0,\pi]/ \mathcal{E}_x,
\end{array} \right.\\
\end{eqnarray}
where $\epsilon\in[0,\pi]$ is the polar angle in the sphere. For the colourings
that we have considered here, $x = 1, 2, 3, 4, 3_\delta$, we define

\begin{eqnarray}
\label{equationC5}
\mathcal{E}_1 &&\equiv \biggl[0,\frac{\pi}{2}\biggr],\nonumber\\
\mathcal{E}_2 &&\equiv
\biggl[0,\frac{\pi}{4}\biggr]\bigcup\biggl[\frac{\pi}{2},\frac{3\pi}{4}\biggr],\nonumber\\
\mathcal{E}_3 &&\equiv
\bigcup_{k=0}^{2}\biggl[k\frac{\pi}{3},(2k+1)\frac{\pi}{6}\biggr],\nonumber\\
\mathcal{E}_4 &&\equiv
\bigcup_{k=0}^{3}\biggl[k\frac{\pi}{4},(2k+1)\frac{\pi}{8}\biggr],\nonumber\\
\mathcal{E}_{3_{\delta}}
&&\equiv\biggl[0,\frac{\pi}{6}+\delta\biggr]\bigcup\biggl[\frac{\pi}{3},\frac{\pi}{2}\biggr]\bigcup\biggl[\frac{2\pi}{3},\frac{5\pi}{6}-\delta\biggr],\nonumber\\
\end{eqnarray}
where $-\frac{\pi}{18}\leq\delta\leq\frac{\pi}{24}$. Notice that colouring
$3_\delta$ reduces
to colouring 3 if $\delta=0$.

\subsection{\label{appendix B}Expressions for the correlations}

We use the azimuthal symmetry of the colourings $x=2,3,4,3_\delta$ defined in
Appendix~\ref{appendix A}, the antipodal property (\ref{eq:antipodalx}) and the
constraint (\ref{eq:17}) to reduce the correlation given by (\ref{eq:14}) to:
\begin{equation}
C_x(\theta)=-\frac{1}{\pi}\int_{0}^{\frac{\pi}{2}}d\epsilon\sin{\epsilon}a_x(\epsilon)\int_{0}^{\pi}d\omega
a_x[\alpha(\theta,\epsilon,\omega)],
\end{equation}
where $\alpha(\theta,\epsilon,\omega)$ is given by Eq.~(\ref{eq:10}). We
computed the integral with respect to $\omega$ in
the previous expression. We define the function
\begin{equation}
\label{eq:b1}
\chi(\theta,a,b,\alpha)\equiv \frac{2}{\pi}\int_a^b
d\epsilon\sin\epsilon\arccos
\Bigl(\frac{\cos\theta\cos\epsilon-\cos\alpha}{\sin\theta\sin\epsilon}\Bigr),
\end{equation}
where $a,b,\alpha\in[0,\pi]$ and $\theta\in\bigl[0,\frac{\pi}{2}\bigr]$. We
obtained the following expressions for the correlations
$C_x(\theta)$:

\begin{eqnarray}
\label{eq:b2}
C_2(\theta)&&=\left\{\begin{array}{rl}
h_2^1(\theta) & \text{if } \theta\in [0,\pi/4],\nonumber\\
h_2^2(\theta) & \text{if } \theta\in (\pi/4,\pi/2],\nonumber\\
\end{array} \right.\\
C_3(\theta)&&=\left\{\begin{array}{rl}
h_3^1(\theta) & \text{if } \theta\in [0,\pi/6],\nonumber\\
h_3^2(\theta) & \text{if } \theta\in (\pi/6,\pi/4],\nonumber\\
h_3^3(\theta) & \text{if } \theta\in (\pi/4,\pi/3],\nonumber\\
h_3^4(\theta) & \text{if } \theta\in (\pi/3,\pi/2],\nonumber\\
\end{array} \right.\\
C_4(\theta)&&=\left\{\begin{array}{rl}
h_4^1(\theta) & \text{if } \theta\in [0,\pi/8],\nonumber\\
h_4^2(\theta) & \text{if } \theta\in (\pi/8,\pi/4],\nonumber\\
h_4^3(\theta) & \text{if } \theta\in (\pi/4,3\pi/8],\nonumber\\
h_4^4(\theta) & \text{if } \theta\in (3\pi/8,\pi/2],\nonumber\\
\end{array} \right.\\
C_{3_\delta}(\theta)&&=\left\{\begin{array}{rl}
r_\delta^1(\theta) & \text{if } \delta\in \bigl[-\frac{\pi}{18},0\bigr] \text{
and } \theta\in \bigl[\frac{\pi}{3},\frac{\pi}{3}- \delta \bigr],\nonumber\\
r_\delta^2(\theta) & \text{if } \delta\in \bigl[-\frac{\pi}{18},0\bigr]  \text{
and } \theta\in \bigl(\frac{\pi}{3}- \delta,\frac{\pi}{2}+\delta \bigr],\nonumber\\
r_\delta^3(\theta) & \text{if }\delta\in \bigl[-\frac{\pi}{18},0\bigr] \text{
and }  \theta\in \bigl(\frac{\pi}{2}+\delta,\frac{\pi}{2}\bigr],\nonumber\\
r_\delta^4(\theta) & \text{if } \delta\in \bigl(0,\frac{\pi}{24}\bigr] \text{
and } \theta\in \bigl[\frac{\pi}{3},\frac{\pi}{3}+2\delta\bigr],\nonumber\\
r_\delta^2(\theta) & \text{if } \delta\in \bigl(0,\frac{\pi}{24}\bigr] \text{
and } \theta\in \bigl(\frac{\pi}{3}+2\delta,\frac{\pi}{2}-\delta\bigr],\nonumber\\
r_\delta^5(\theta) & \text{if } \delta\in \bigl(0,\frac{\pi}{24}\bigr] \text{
and } \theta\in \bigl(\frac{\pi}{2}-\delta,\frac{\pi}{2}\bigr],\nonumber\\
\end{array} \right.\nonumber\\
\end{eqnarray}
where

\begin{widetext}
\begin{eqnarray}
h_2^1(\theta)&&\equiv
-1+2\biggl[\cos\Bigl(\frac{\pi}{4}\Bigr)-\cos\Bigl(\frac{\pi}{4}+\theta\Bigr)\biggr]+\chi\Bigl(\theta,\frac{\pi}{4}-\theta,\frac{\pi}{4},\frac{\pi}{4}\Bigr)-\chi\Bigl(\theta,\frac{\pi}{4},\frac{\pi}{4}+\theta,\frac{\pi}{4}\Bigr)+\chi\Bigl(\theta,\frac{\pi}{2}-\theta,\frac{\pi}{2},\frac{\pi}{2}\Bigr),\nonumber\\
h_2^2(\theta)&&\equiv
1+2\biggl[\cos\Bigl(\frac{\pi}{4}\Bigr)-\cos\Bigl(\theta-\frac{\pi}{4}\Bigr)\biggr]+\chi\Bigl(\theta,\theta-\frac{\pi}{4},\frac{\pi}{4},\frac{\pi}{4}\Bigr)-\chi\Bigl(\theta,\frac{\pi}{2}-\theta,\frac{\pi}{4},\frac{\pi}{2}\Bigr)+\chi\Bigl(\theta,\frac{\pi}{4},\frac{\pi}{2},\frac{\pi}{2}\Bigr)\nonumber\\
&&\qquad-\chi\Bigl(\theta,\frac{\pi}{4},\frac{\pi}{2},\frac{\pi}{4}\Bigr)-\chi\Bigl(\theta,\frac{3\pi}{4}-\theta,\frac{\pi}{2},\frac{3\pi}{4}\Bigr);\nonumber
\end{eqnarray}

\begin{eqnarray}
h_3^1(\theta)&&\equiv
-1+2\biggl[\cos\Bigl(\frac{\pi}{6}\Bigr)-\cos\Bigl(\frac{\pi}{6}+\theta\Bigr)+\cos\Bigl(\frac{\pi}{3}\Bigr)-\cos\Bigl(\frac{\pi}{3}+\theta\Bigr)\Bigr]+\chi\Bigl(\theta,\frac{\pi}{6}-\theta,\frac{\pi}{6},\frac{\pi}{6}\Bigr)
-\chi\Bigl(\theta,\frac{\pi}{6},\frac{\pi}{6}+\theta,\frac{\pi}{6}\Bigr)\nonumber\\
&&\qquad+\chi\Bigl(\theta,\frac{\pi}{3}-\theta,\frac{\pi}{3},\frac{\pi}{3}\Bigr)-\chi\Bigl(\theta,\frac{\pi}{3},\frac{\pi}{3}+\theta,\frac{\pi}{3}\Bigr)+\chi\Bigl(\theta,\frac{\pi}{2}-\theta,\frac{\pi}{2},\frac{\pi}{2}\Bigr),\nonumber\\
h_3^2(\theta)&&\equiv
1+2\biggl[\cos\Bigl(\frac{\pi}{6}\Bigr)-\cos\Bigl(\theta-\frac{\pi}{6}\Bigr)+\cos\Bigl(\frac{\pi}{6}+\theta\Bigr)-\cos\Bigl(\frac{\pi}{3}\Bigr)\Bigr]+\chi\Bigl(\theta,\theta-\frac{\pi}{6},\frac{\pi}{6},\frac{\pi}{6}\Bigr)
-\chi\Bigl(\theta,\frac{\pi}{3}-\theta,\frac{\pi}{6},\frac{\pi}{3}\Bigr)\nonumber\\
&&\qquad+\chi\Bigl(\theta,\frac{\pi}{6},\frac{\pi}{2}-\theta,\frac{\pi}{3}\Bigr)-\chi\Bigl(\theta,\frac{\pi}{6},\frac{\pi}{3},\frac{\pi}{6}\Bigr)+\chi\Bigl(\theta,\frac{\pi}{2}-\theta,\frac{\pi}{3},\frac{\pi}{3}\Bigr)-\chi\Bigl(\theta,\frac{\pi}{2}-\theta,\frac{\pi}{3},\frac{\pi}{2}\Bigr)+\chi\Bigl(\theta,\frac{\pi}{3},\frac{\pi}{6}+\theta,\frac{\pi}{6}\Bigr)\nonumber\\
&&\qquad+\chi\Bigl(\theta,\frac{\pi}{3},\frac{\pi}{2},\frac{\pi}{2}\Bigr)-\chi\Bigl(\theta,\frac{\pi}{3},\frac{\pi}{2},\frac{\pi}{3}\Bigr)-\chi\Bigl(\theta,\frac{2\pi}{3}-\theta,\frac{\pi}{2},\frac{2\pi}{3}\Bigr),\nonumber\\
h_3^3(\theta)&&\equiv
1+2\biggl[\cos\Bigl(\frac{\pi}{6}\Bigr)-\cos\Bigl(\theta-\frac{\pi}{6}\Bigr)+\cos\Bigl(\frac{\pi}{6}+\theta\Bigr)-\cos\Bigl(\frac{\pi}{3}\Bigr)\Bigr]-\chi\Bigl(\theta,\frac{\pi}{3}-\theta,\frac{\pi}{6},\frac{\pi}{3}\Bigr)+\chi\Bigl(\theta,\theta-\frac{\pi}{6},\frac{\pi}{6},\frac{\pi}{6}\Bigr)\nonumber\\
&&\qquad+\chi\Bigl(\theta,\frac{\pi}{6},\frac{\pi}{3},\frac{\pi}{3}\Bigr)-\chi\Bigl(\theta,\frac{\pi}{6},\frac{\pi}{3},\frac{\pi}{6}\Bigr)-\chi\Bigl(\theta,\frac{\pi}{2}-\theta,\frac{\pi}{3},\frac{\pi}{2}\Bigr)+\chi\Bigl(\theta,\frac{\pi}{3},\frac{\pi}{2},\frac{\pi}{2}\Bigr)-\chi\Bigl(\theta,\frac{\pi}{3},\frac{\pi}{2},\frac{\pi}{3}\Bigr)\nonumber\\
&&\qquad+\chi\Bigl(\theta,\frac{\pi}{3},\frac{\pi}{6}+\theta,\frac{\pi}{6}\Bigr)-\chi\Bigl(\theta,\frac{2\pi}{3}-\theta,\frac{\pi}{2},\frac{2\pi}{3}\Bigr),\nonumber\\
h_3^4(\theta)&&\equiv
-1+2\biggl[\cos\Bigl(\theta-\frac{\pi}{3}\Bigr)-\cos\Bigl(\frac{\pi}{6}\Bigr)+\cos\Bigl(\theta-\frac{\pi}{6}\Bigr)-\cos\Bigl(\frac{\pi}{3}\Bigr)\Bigr]
-\chi\Bigl(\theta,\theta-\frac{\pi}{3},\frac{\pi}{6},\frac{\pi}{3}\Bigr)+\chi\Bigl(\theta,\frac{\pi}{2}-\theta,\frac{\pi}{6},\frac{\pi}{2}\Bigr)\nonumber\\
&&\qquad+\chi\Bigl(\theta,\frac{\pi}{6},\frac{\pi}{3},\frac{\pi}{3}\Bigr)-\chi\Bigl(\theta,\frac{\pi}{6},\frac{\pi}{3},\frac{\pi}{2}\Bigr)-\chi\Bigl(\theta,\theta-\frac{\pi}{6},\frac{\pi}{3},\frac{\pi}{6}\Bigr)+\chi\Bigl(\theta,\frac{2\pi}{3}-\theta,\frac{\pi}{3},\frac{2\pi}{3}\Bigr)-\chi\Bigl(\theta,\frac{\pi}{3},\frac{\pi}{2},\frac{2\pi}{3}\Bigr)\nonumber\\
&&\qquad+\chi\Bigl(\theta,\frac{\pi}{3},\frac{\pi}{2},\frac{\pi}{2}\Bigr)-\chi\Bigl(\theta,\frac{\pi}{3},\frac{\pi}{2},\frac{\pi}{3}\Bigr)+\chi\Bigl(\theta,\frac{\pi}{3},\frac{\pi}{2},\frac{\pi}{6}\Bigr)+\chi\Bigl(\theta,\frac{5\pi}{6}-\theta,\frac{\pi}{2},\frac{5\pi}{6}\Bigr);\nonumber
\end{eqnarray}

\begin{eqnarray}
h_4^1(\theta)&&\equiv -1+2
\biggl[\cos\Bigl(\frac{\pi}{8}\Bigr)-\cos\Bigl(\frac{\pi}{8}+\theta\Bigr)+\cos\Bigl(\frac{\pi}{4}\Bigr)-\cos\Bigl(\frac{\pi}{4}+\theta\Bigr)+\cos\Bigl(\frac{3\pi}{8}\Bigr)-\cos\Bigl(\frac{3\pi}{8}+\theta\Bigr)\biggr]+\chi\Bigl(\theta,\frac{\pi}{8}-\theta,\frac{\pi}{8},\frac{\pi}{8}\Bigr)\nonumber\\
&&\qquad-\chi\Bigl(\theta,\frac{\pi}{8},\frac{\pi}{8}+\theta,\frac{\pi}{8}\Bigr)+\chi\Bigl(\theta,\frac{\pi}{4}-\theta,\frac{\pi}{4},\frac{\pi}{4}\Bigr)-\chi\Bigl(\theta,\frac{\pi}{4},\frac{\pi}{4}+\theta,\frac{\pi}{4}\Bigr)+\chi\Bigl(\theta,\frac{3\pi}{8}-\theta,\frac{3\pi}{8},\frac{3\pi}{8}\Bigr)\nonumber\\
&&\qquad-\chi\Bigl(\theta,\frac{3\pi}{8},\frac{3\pi}{8}+\theta,\frac{3\pi}{8}\Bigr)+\chi\Bigl(\theta,\frac{\pi}{2}-\theta,\frac{\pi}{2},\frac{\pi}{2}\Bigr),\nonumber\\
h_4^2(\theta)&&\equiv
1+2\biggl[\cos\Bigl(\frac{\pi}{8}\Bigr)-\cos\Bigl(\theta-\frac{\pi}{8}\Bigr)+\cos\Bigl(\theta+\frac{\pi}{8}\Bigr)-\cos\Bigl(\frac{\pi}{4}\Bigr)+\cos\Bigl(\theta+\frac{\pi}{4}\Bigr)-\cos\Bigl(\frac{3\pi}{8}\Bigr)\biggr]+\chi\Bigl(\theta,\theta-\frac{\pi}{8},\frac{\pi}{8},\frac{\pi}{8}\Bigr)\nonumber\\
&&\qquad-\chi\Bigl(\theta,\frac{\pi}{4}-\theta,\frac{\pi}{8},\frac{\pi}{4}\Bigr)+\chi\Bigl(\theta,\frac{\pi}{8},\frac{\pi}{4},\frac{\pi}{4}\Bigr)-\chi\Bigl(\theta,\frac{\pi}{8},\frac{\pi}{4},\frac{\pi}{8}\Bigr)-\chi\Bigl(\theta,\frac{3\pi}{8}-\theta,\frac{\pi}{4},\frac{3\pi}{8}\Bigr)+\chi\Bigl(\theta,\frac{\pi}{4},\frac{\pi}{8}+\theta,\frac{\pi}{8}\Bigr)\nonumber\\
&&\qquad+\chi\Bigl(\theta,\frac{\pi}{4},\frac{3\pi}{8},\frac{3\pi}{8}\Bigr)-\chi\Bigl(\theta,\frac{\pi}{4},\frac{3\pi}{8},\frac{\pi}{4}\Bigr)-\chi\Bigl(\theta,\frac{\pi}{2}-\theta,\frac{3\pi}{8},\frac{\pi}{2}\Bigr)+\chi\Bigl(\theta,\frac{3\pi}{8},\frac{\pi}{4}+\theta,\frac{\pi}{4}\Bigr)+\chi\Bigl(\theta,\frac{3\pi}{8},\frac{\pi}{2},\frac{\pi}{2}\Bigr)\nonumber\\
&&\qquad-\chi\Bigl(\theta,\frac{3\pi}{8},\frac{\pi}{2},\frac{3\pi}{8}\Bigr)-\chi\Bigl(\theta,\frac{5\pi}{8}-\theta,\frac{\pi}{2},\frac{5\pi}{8}\Bigr),\nonumber\\
h_4^3(\theta)&&\equiv
-1+2\biggl[\cos\Bigl(\theta-\frac{\pi}{4}\Bigr)-\cos\Bigl(\frac{\pi}{8}\Bigr)+\cos\Bigl(\theta-\frac{\pi}{8}\Bigr)-\cos\Bigl(\frac{\pi}{4}\Bigr)+\cos\Bigl(\frac{3\pi}{8}\Bigr)-\cos\Bigl(\theta+\frac{\pi}{8}\Bigr)\biggr]-\chi\Bigl(\theta,\theta-\frac{\pi}{4},\frac{\pi}{8},\frac{\pi}{4}\Bigr)\nonumber\\
&&\qquad+\chi\Bigl(\theta,\frac{3\pi}{8}-\theta,\frac{\pi}{8},\frac{3\pi}{8}\Bigr)-\chi\Bigl(\theta,\frac{\pi}{8},\frac{\pi}{4},\frac{3\pi}{8}\Bigr)+\chi\Bigl(\theta,\frac{\pi}{8},\frac{\pi}{4},\frac{\pi}{4}\Bigr)-\chi\Bigl(\theta,\theta-\frac{\pi}{8},\frac{\pi}{4},\frac{\pi}{8}\Bigr)+\chi\Bigl(\theta,\frac{\pi}{2}-\theta,\frac{\pi}{4},\frac{\pi}{2}\Bigr)\nonumber\\
&&\qquad-\chi\Bigl(\theta,\frac{\pi}{4},\frac{3\pi}{8},\frac{\pi}{2}\Bigr)+\chi\Bigl(\theta,\frac{\pi}{4},\frac{3\pi}{8},\frac{3\pi}{8}\Bigr)-\chi\Bigl(\theta,\frac{\pi}{4},\frac{3\pi}{8},\frac{\pi}{4}\Bigr)+\chi\Bigl(\theta,\frac{\pi}{4},\frac{3\pi}{8},\frac{\pi}{8}\Bigr)+\chi\Bigl(\theta,\frac{5\pi}{8}-\theta,\frac{3\pi}{8},\frac{5\pi}{8}\Bigr)\nonumber\\
&&\qquad-\chi\Bigl(\theta,\frac{3\pi}{8},\frac{\pi}{2},\frac{5\pi}{8}\Bigr)+\chi\Bigl(\theta,\frac{3\pi}{8},\frac{\pi}{2},\frac{\pi}{2}\Bigr)-\chi\Bigl(\theta,\frac{3\pi}{8},\frac{\pi}{2},\frac{3\pi}{8}\Bigr)+\chi\Bigl(\theta,\frac{3\pi}{8},\frac{\pi}{2},\frac{\pi}{4}\Bigr)-\chi\Bigl(\theta,\frac{3\pi}{8},\frac{\pi}{8}+\theta,\frac{\pi}{8}\Bigr)\nonumber\\
&&\qquad+\chi\Bigl(\theta,\frac{3\pi}{4}-\theta,\frac{\pi}{2},\frac{3\pi}{4}\Bigr),\nonumber\\
h_4^4(\theta)&&\equiv
1+2\biggl[\cos\Bigl(\frac{\pi}{8}\Bigr)-\cos\Bigl(\theta-\frac{3\pi}{8}\Bigr)+\cos\Bigl(\frac{\pi}{4}\Bigr)-\cos\Bigl(\theta-\frac{\pi}{4}\Bigr)+\cos\Bigl(\frac{3\pi}{8}\Bigr)-\cos\Bigl(\theta-\frac{\pi}{8}\Bigr)\biggr]+\chi\Bigl(\theta,\theta-\frac{3\pi}{8},\frac{\pi}{8},\frac{3\pi}{8}\Bigr)\nonumber\\
&&\qquad
-\chi\Bigl(\theta,\frac{\pi}{2}-\theta,\frac{\pi}{8},\frac{\pi}{2}\Bigr)+\chi\Bigl(\theta,\frac{\pi}{8},\frac{\pi}{4},\frac{\pi}{2}\Bigr)-\chi\Bigl(\theta,\frac{\pi}{8},\frac{\pi}{4},\frac{3\pi}{8}\Bigr)+\chi\Bigl(\theta,\theta-\frac{\pi}{4},\frac{\pi}{4},\frac{\pi}{4}\Bigr)-\chi\Bigl(\theta,\frac{5\pi}{8}-\theta,\frac{\pi}{4},\frac{5\pi}{8}\Bigr)\nonumber\\
&&\qquad+\chi\Bigl(\theta,\frac{\pi}{4},\frac{3\pi}{8},\frac{5\pi}{8}\Bigr)-\chi\Bigl(\theta,\frac{\pi}{4},\frac{3\pi}{8},\frac{\pi}{2}\Bigr)+\chi\Bigl(\theta,\frac{\pi}{4},\frac{3\pi}{8},\frac{3\pi}{8}\Bigr)-\chi\Bigl(\theta,\frac{\pi}{4},\frac{3\pi}{8},\frac{\pi}{4}\Bigr)+\chi\Bigl(\theta,\theta-\frac{\pi}{8},\frac{3\pi}{8},\frac{\pi}{8}\Bigr)\nonumber\\
&&\qquad-\chi\Bigl(\theta,\frac{3\pi}{4}-\theta,\frac{3\pi}{8},\frac{3\pi}{4}\Bigr)+\chi\Bigl(\theta,\frac{3\pi}{8},\frac{\pi}{2},\frac{3\pi}{4}\Bigr)-\chi\Bigl(\theta,\frac{3\pi}{8},\frac{\pi}{2},\frac{5\pi}{8}\Bigr)+\chi\Bigl(\theta,\frac{3\pi}{8},\frac{\pi}{2},\frac{\pi}{2}\Bigr)-\chi\Bigl(\theta,\frac{3\pi}{8},\frac{\pi}{2},\frac{3\pi}{8}\Bigr)\nonumber\\
&&\qquad+\chi\Bigl(\theta,\frac{3\pi}{8},\frac{\pi}{2},\frac{\pi}{4}\Bigr)-\chi\Bigl(\theta,\frac{3\pi}{8},\frac{\pi}{2},\frac{\pi}{8}\Bigr)-\chi\Bigl(\theta,\frac{7\pi}{8}-\theta,\frac{\pi}{2},\frac{7\pi}{8}\Bigr);\nonumber
\end{eqnarray}

\begin{eqnarray}
r_\delta^1(\theta)&&\equiv
-1+2\biggl[\cos\Bigl(\theta-\frac{\pi}{3}\Bigr)-\cos\Bigl(\frac{\pi}{6}+\delta\Bigr)+\cos\Bigl(\theta-\frac{\pi}{6}-\delta\Bigr)-\cos\Bigl(\frac{\pi}{3}\Bigr)+\cos\Bigl(\theta+\frac{\pi}{6}+\delta\Bigr)\biggr]\nonumber\\
&&\qquad-\chi\Bigl(\theta,\theta-\frac{\pi}{3},\frac{\pi}{6}+\delta,\frac{\pi}{3}\Bigr)+\chi\Bigl(\theta,\frac{\pi}{6}+\delta,\frac{\pi}{3},\frac{\pi}{3}\Bigr)-\chi\Bigl(\theta,\frac{\pi}{2}-\theta,\frac{\pi}{3},\frac{\pi}{2}\Bigr)-\chi\Bigl(\theta,\theta-\frac{\pi}{6}-\delta,\frac{\pi}{3},\frac{\pi}{6}+\delta\Bigr)\nonumber\\
&&\qquad+\chi\Bigl(\theta,\frac{2\pi}{3}-\theta,\frac{\pi}{3},\frac{2\pi}{3}\Bigr)
-\chi\Bigl(\theta,\frac{\pi}{3},\frac{\pi}{2},\frac{2\pi}{3}\Bigr)+\chi\Bigl(\theta,\frac{\pi}{3},\frac{\pi}{2},\frac{\pi}{2}\Bigr)-\chi\Bigl(\theta,\frac{\pi}{3},\frac{\pi}{2},\frac{\pi}{3}\Bigr)+\chi\Bigl(\theta,\frac{\pi}{3},\theta+\frac{\pi}{6}+\delta,\frac{\pi}{6}+\delta\Bigr),\nonumber\\
r_\delta^2(\theta)&&\equiv
-1+2\biggl[\cos\Bigl(\theta-\frac{\pi}{3}\Bigr)-\cos\Bigl(\frac{\pi}{6}+\delta\Bigr)+\cos\Bigl(\theta-\frac{\pi}{6}-\delta\Bigr)-\cos\Bigl(\frac{\pi}{3}\Bigr)\biggr]-\chi\Bigl(\theta,\theta-\frac{\pi}{3},\frac{\pi}{6}+\delta,\frac{\pi}{3}\Bigr)\nonumber\\
&&\qquad+\chi\Bigl(\theta,\frac{\pi}{2}-\theta,\frac{\pi}{6}+\delta,\frac{\pi}{2}\Bigr)-\chi\Bigl(\theta,\frac{\pi}{6}+\delta,\frac{\pi}{3},\frac{\pi}{2}\Bigr)+\chi\Bigl(\theta,\frac{\pi}{6}+\delta,\frac{\pi}{3},\frac{\pi}{3}\Bigr)-\chi\Bigl(\theta,\theta-\frac{\pi}{6}-\delta,\frac{\pi}{3},\frac{\pi}{6}+\delta\Bigr)\nonumber\\
&&\qquad
+\chi\Bigl(\theta,\frac{2\pi}{3}-\theta,\frac{\pi}{3},\frac{2\pi}{3}\Bigr)-\chi\Bigl(\theta,\frac{\pi}{3},\frac{\pi}{2},\frac{2\pi}{3}\Bigr)+\chi\Bigl(\theta,\frac{\pi}{3},\frac{\pi}{2},\frac{\pi}{2}\Bigr)-\chi\Bigl(\theta,\frac{\pi}{3},\frac{\pi}{2},\frac{\pi}{3}\Bigr)+\chi\Bigl(\theta,\frac{\pi}{3},\frac{\pi}{2},\frac{\pi}{6}+\delta\Bigr)\nonumber\\
&&\qquad+\chi\Bigl(\theta,\frac{5\pi}{6}-\delta-\theta,\frac{\pi}{2},\frac{5\pi}{6}-\delta\Bigr),\nonumber\\
r_\delta^3(\theta)&&\equiv
-1+2\biggl[\cos\Bigl(\frac{\pi}{6}+\delta\Bigr)-\cos\Bigl(\theta-\frac{\pi}{3}\Bigr)+\cos\Bigl(\frac{\pi}{3}\Bigr)-\cos\Bigl(\theta-\frac{\pi}{6}-\delta\Bigr)\biggr]+\chi\Bigl(\theta,\frac{\pi}{2}-\theta,\frac{\pi}{6}+\delta,\frac{\pi}{2}\Bigr)\nonumber\\
&&\qquad-\chi\Bigl(\theta,\frac{\pi}{6}+\delta,\frac{\pi}{3},\frac{\pi}{2}\Bigr)+\chi\Bigl(\theta,\theta-\frac{\pi}{3},\frac{\pi}{3},\frac{\pi}{3}\Bigr)+\chi\Bigl(\theta,\frac{2\pi}{3}-\theta,\frac{\pi}{3},\frac{2\pi}{3}\Bigr)-\chi\Bigl(\theta,\frac{\pi}{3},\frac{\pi}{2},\frac{2\pi}{3}\Bigr)+\chi\Bigl(\theta,\frac{\pi}{3},\frac{\pi}{2},\frac{\pi}{2}\Bigr)\nonumber\\
&&\qquad-\chi\Bigl(\theta,\frac{\pi}{3},\frac{\pi}{2},\frac{\pi}{3}\Bigr)+\chi\Bigl(\theta,\theta-\frac{\pi}{6}-\delta,\frac{\pi}{2},\frac{\pi}{6}+\delta\Bigr)+\chi\Bigl(\theta,\frac{5\pi}{6}-\delta-\theta,\frac{\pi}{2},\frac{5\pi}{6}-\delta\Bigr),\nonumber\\
r_\delta^4(\theta)&&\equiv
-1+2\biggl[\cos\Bigl(\theta-\frac{\pi}{3}\Bigr)-\cos\Bigl(\theta-\frac{\pi}{6}-\delta\Bigr)+\cos\Bigl(\frac{\pi}{6}+\delta\Bigr)-\cos\Bigl(\frac{\pi}{3}\Bigr)\biggr]-\chi\Bigl(\theta,\theta-\frac{\pi}{3},\frac{\pi}{6}+\delta,\frac{\pi}{3}\Bigr)\nonumber\\
&&\qquad+\chi\Bigl(\theta,\theta-\frac{\pi}{6}-\delta,\frac{\pi}{6}+\delta,\frac{\pi}{6}+\delta\Bigr)+\chi\Bigl(\theta,\frac{\pi}{2}-\theta,\frac{\pi}{6}+\delta,\frac{\pi}{2}\Bigr)-\chi\Bigl(\theta,\frac{\pi}{6}+\delta,\frac{\pi}{3},\frac{\pi}{2}\Bigr)+\chi\Bigl(\theta,\frac{\pi}{6}+\delta,\frac{\pi}{3},\frac{\pi}{3}\Bigr)\nonumber\\
&&\qquad-\chi\Bigl(\theta,\frac{\pi}{6}+\delta,\frac{\pi}{3},\frac{\pi}{6}+\delta\Bigr)+\chi\Bigl(\theta,\frac{2\pi}{3}-\theta,\frac{\pi}{3},\frac{2\pi}{3}\Bigr)-\chi\Bigl(\theta,\frac{\pi}{3},\frac{\pi}{2},\frac{2\pi}{3}\Bigr)+\chi\Bigl(\theta,\frac{\pi}{3},\frac{\pi}{2},\frac{\pi}{2}\Bigr)-\chi\Bigl(\theta,\frac{\pi}{3},\frac{\pi}{2},\frac{\pi}{3}\Bigr)\nonumber\\
&&\qquad+\chi\Bigl(\theta,\frac{\pi}{3},\frac{\pi}{2},\frac{\pi}{6}+\delta\Bigr)+\chi\Bigl(\theta,\frac{5\pi}{6}-\delta-\theta,\frac{\pi}{2},\frac{5\pi}{6}-\delta\Bigr),\nonumber\\
r_\delta^5(\theta)&&\equiv
-1+2\biggl[\cos\Bigl(\theta-\frac{\pi}{3}\Bigr)-\cos\Bigl(\frac{\pi}{6}+\delta\Bigr)+\cos\Bigl(\theta-\frac{\pi}{6}-\delta\Bigr)-\cos\Bigl(\frac{\pi}{3}\Bigr)\biggr]+\chi\Bigl(\theta,\frac{\pi}{2}-\theta,\frac{\pi}{6}+\delta,\frac{\pi}{2}\Bigr)\nonumber\\
&&\qquad-\chi\Bigl(\theta,\theta-\frac{\pi}{3},\frac{\pi}{6}+\delta,\frac{\pi}{3}\Bigr)-\chi\Bigl(\theta,\frac{2\pi}{3}-\theta,\frac{\pi}{6}+\delta,\frac{2\pi}{3}\Bigr)+\chi\Bigl(\theta,\frac{\pi}{6}+\delta,\frac{\pi}{3},\frac{2\pi}{3}\Bigr)-\chi\Bigl(\theta,\frac{\pi}{6}+\delta,\frac{\pi}{3},\frac{\pi}{2}\Bigr)\nonumber\\
&&\qquad+\chi\Bigl(\theta,\frac{\pi}{6}+\delta,\frac{\pi}{3},\frac{\pi}{3}\Bigr)-\chi\Bigl(\theta,\theta-\frac{\pi}{6}-\delta,\frac{\pi}{3},\frac{\pi}{6}+\delta\Bigr)-\chi\Bigl(\theta,\frac{5\pi}{6}-\delta-\theta,\frac{\pi}{3},\frac{5\pi}{6}-\delta\Bigr)+\chi\Bigl(\theta,\frac{\pi}{3},\frac{\pi}{2},\frac{5\pi}{6}-\delta\Bigr)\nonumber\\
&&\qquad-\chi\Bigl(\theta,\frac{\pi}{3},\frac{\pi}{2},\frac{2\pi}{3}\Bigr)+\chi\Bigl(\theta,\frac{\pi}{3},\frac{\pi}{2},\frac{\pi}{2}\Bigr)-\chi\Bigl(\theta,\frac{\pi}{3},\frac{\pi}{2},\frac{\pi}{3}\Bigr)+\chi\Bigl(\theta,\frac{\pi}{3},\frac{\pi}{2},\frac{\pi}{6}+\delta\Bigr).\nonumber
\end{eqnarray}
\end{widetext}

\subsection{\label{appendix C}Proof of Equation~(\ref{eq:22})}

Let $0\leq\tau\ll 1$. To show Eq.~(\ref{eq:22}), we expand $C_3\bigl(\frac{\pi}{2}-\tau\bigr)$ in its Taylor series to obtain
\begin{equation}
\label{eq:c2}
C_3\Bigl(\frac{\pi}{2}-\tau\Bigr)=C_3\Bigl(\frac{\pi}{2}\Bigr)+\tau\Bigl[\frac{d}{d\tau}C_3\Bigl(\frac{\pi}{2}-\tau\Bigr)\Bigr]_{\tau=0}+\mathcal{O}(\tau^2).
\end{equation}
As shown in the main text, the correlation satisfies $C_x(\frac{\pi}{2})=0$ for
any pair of colourings labelled by $x$ that we consider. Thus, we have that
$C_3\bigl(\frac{\pi}{2}\bigr)=0$. From
Eq.~(\ref{eq:b2}), we have that $C_3\bigl(\frac{\pi}{2}-\tau\bigr) =
h_3^4\bigl(\frac{\pi}{2}-\tau\bigr)$  for $0\leq\tau\ll 1$. Thus, we only need
to show that
\begin{equation}
\label{eq:c3}
\Bigl[\frac{d}{d\theta}h_3^4(\theta)\Bigr]_{\theta=\pi/2}=1.5.
\end{equation}

The function $h_3^4(\theta)$ has terms of the form
\begin{equation}
\label{eq:c4}
\chi(\theta,a,b,\alpha)\equiv\int_{a}^{b}d\epsilon\xi(\theta,\epsilon,\alpha),
\end{equation}
where
\begin{equation}
\label{eq:c5}
\xi(\theta,\epsilon,\alpha)\equiv \frac{2}{\pi}\sin\epsilon\arccos
\Bigl(\frac{\cos\theta\cos\epsilon-\cos\alpha}{\sin\theta\sin\epsilon}\Bigr),
\end{equation}
as defined by Eq.~(\ref{eq:b1}). Differentiating the function $\chi$, we obtain
\begin{eqnarray}
\label{eq:c6}
\frac{d}{d\theta}\chi(\theta,a,b,\alpha)&&=\xi(\theta,b,\alpha)\frac{db}{d\theta}-\xi(\theta,a,\alpha)\frac{da}{d\theta}\nonumber\\
&&\qquad+\int_{a}^{b}d\epsilon\frac{\partial}{\partial
\theta}\xi(\theta,\epsilon,\alpha).
\end{eqnarray}
We have that
\begin{equation}
\label{eq:c7}
\biggl[\frac{\partial}{\partial
\theta}\xi(\theta,\epsilon,\alpha)\biggr]_{\theta=\pi/2}=\frac{2\cos\epsilon}{\pi\sqrt{1-\bigl(\frac{\cos\alpha}{\sin\epsilon}\bigr)^2}}.
\end{equation}
We obtain that
\begin{equation}
\label{eq:c8}
\frac{2}{\pi}\int\limits_{a}^{b}\frac{d\epsilon\cos\epsilon}{\sqrt{1-\bigl(\frac{\cos\alpha}{\sin\epsilon}\bigr)^2}}=\mu(a,b,\alpha),
\end{equation}
where
\begin{eqnarray}
\label{eq:c9}
\mu(a,b,\alpha) && \equiv \frac{2}{\pi}\Bigl(\sqrt{\sin^2
b-\cos^2\alpha}-\sqrt{\sin^2 a-\cos^2\alpha}\Bigr)\nonumber\\
\end{eqnarray}
for $\cos^2\alpha\leq \sin^2 b$ and $\cos^2\alpha\leq \sin^2 a$.
We define
\begin{equation}
\label{eq:c10}
\nu(\epsilon,\alpha)\equiv \xi\Bigl(\frac{\pi}{2},\epsilon,\alpha\Bigr).
\end{equation}
From the definition of $h_3^4(\theta)$ given in Appendix \ref{appendix B} and
Eqs.~(\ref{eq:c6}) -- (\ref{eq:c10}), it is straightforward to obtain that
\begin{widetext}
\begin{eqnarray}
\Bigl[\frac{d}{d\theta}h_3^4(\theta)\Bigr]_{\theta=\pi/2}&&=-2\Bigl[\sin\Bigl(\frac{\pi}{6}\Bigr)+\sin\Bigl(\frac{\pi}{3}\Bigr)\Bigr]+\nu\Bigl(0,\frac{\pi}{2}\Bigr)+\nu\Bigl(\frac{\pi}{6},\frac{\pi}{3}\Bigr)+\nu\Bigl(\frac{\pi}{6},\frac{2\pi}{3}\Bigr)+\nu\Bigl(\frac{\pi}{3},\frac{\pi}{6}\Bigr)+\nu\Bigl(\frac{\pi}{3},\frac{5\pi}{6}\Bigr)\nonumber\\
&&\qquad+\mu\Bigl(0,\frac{\pi}{6},\frac{\pi}{2}\Bigr)-\mu\Bigl(\frac{\pi}{6},\frac{\pi}{6},\frac{\pi}{3}\Bigr)+\mu\Bigl(\frac{\pi}{6},\frac{\pi}{3},\frac{\pi}{3}\Bigr)-\mu\Bigl(\frac{\pi}{6},\frac{\pi}{3},\frac{\pi}{2}\Bigr)+\mu\Bigl(\frac{\pi}{6},\frac{\pi}{3},\frac{2\pi}{3}\Bigr)-\mu\Bigl(\frac{\pi}{3},\frac{\pi}{3},\frac{\pi}{6}\Bigr)\nonumber\\
&&\qquad+\mu\Bigl(\frac{\pi}{3},\frac{\pi}{2},\frac{\pi}{2}\Bigr)+\mu\Bigl(\frac{\pi}{3},\frac{\pi}{2},\frac{\pi}{6}\Bigr)-\mu\Bigl(\frac{\pi}{3},\frac{\pi}{2},\frac{2\pi}{3}\Bigr)-\mu\Bigl(\frac{\pi}{3},\frac{\pi}{2},\frac{\pi}{3}\Bigr)+\mu\Bigl(\frac{\pi}{3},\frac{\pi}{2},\frac{5\pi}{6}\Bigr).
\end{eqnarray}
\end{widetext}
We use Eqs.~(\ref{eq:c5}), (\ref{eq:c9}) and (\ref{eq:c10}), and notice that
$\nu\bigl(0,\frac{\pi}{2}\bigr)=0$ in order to evaluate the previous
expression. We obtain
\begin{equation}
\Bigl[\frac{d}{d\theta}h_3^4(\theta)\Bigr]_{\theta=\pi/2} =
\frac{1}{\pi}\Bigl[6-4\bigl(\sqrt{3}-\sqrt{2}\!~\bigr)\Bigr]=1.5,
\end{equation}
as claimed.

%

\end{document}